\documentclass[aps,prb,twocolumn,showpacs,longbibliography,superscriptaddress]{revtex4-2}
\usepackage{url}
\usepackage{fancyhdr}
\usepackage{amstext, amsmath, amssymb, amsthm}
\usepackage{braket}
\usepackage[margin=2.5cm]{geometry}
\usepackage{array,xcolor,graphicx,url}
\usepackage{verbatim}
\usepackage{float}

\DeclareFixedFont{\ttb}{T1}{txtt}{bx}{n}{12}
\DeclareFixedFont{\ttm}{T1}{txtt}{m}{n}{12}

\usepackage{color}
\definecolor{deepblue}{rgb}{0,0,0.5}
\definecolor{deepred}{rgb}{0.6,0,0}
\definecolor{deepgreen}{rgb}{0,0.5,0}

\usepackage{listings}

\newcommand\pythonstyle{\lstset{
		language=Python,
		basicstyle=\ttm,
		keywordstyle=\ttb\color{deepblue},
		breaklines=true,
		breakatwhitespace=false,
		emphstyle=\ttb\color{deepred},
		stringstyle=\color{deepgreen},
		frame=tb,                         
		showstringspaces=false
}}

\newcommand{\Renyi}[0]{R\'{e}nyi~}

\usepackage[unicode=true]{hyperref}

\newcommand{\be}{\begin{equation}}
\newcommand{\ee}{\end{equation}}
\newcommand{\bea}{\begin{eqnarray}}
\newcommand{\eea}{\end{eqnarray}}

\newcommand{\Tr}[1]{\mathrm{Tr} #1}

\newcommand{\calh}{\mathcal{H}}
\newcommand{\calq}{\mathcal{Q}}
\newcommand{\intz}{\mathbb{Z}}

\begin{document}

\title{Observing Floquet topological order by symmetry resolution}
\author{Daniel Azses}
\affiliation{School of Physics and Astronomy, Tel Aviv University, Tel Aviv 6997801, Israel}

\author{Emanuele G. Dalla Torre}
\affiliation{Department of Physics, Bar-Ilan University, Ramat Gan 5290002, Israel}
\affiliation{Center for Quantum Entanglement Science and Technology, Bar-Ilan University, Ramat Gan 5290002, Israel}

\author{Eran Sela}
\affiliation{School of Physics and Astronomy, Tel Aviv University, Tel Aviv 6997801, Israel}

\begin{abstract}
	Symmetry protected topological order in one dimension leads to protected degeneracies between symmetry blocks of the reduced density matrix. In the presence of periodic driving, topological Floquet phases can be identified in terms of a cycling of these symmetry blocks between different charge quantum numbers. We discuss an example of this phenomenon with an Ising $\intz_2$ symmetry, using both analytic methods and real quantum computers. By adiabatically moving along the phase diagram, we demonstrate that the cycling periodicity is broken in Floquet topological phase transitions. An equivalent signature of the topological Floquet phase is identified as a computational power allowing to teleport quantum information.
\end{abstract}

\maketitle

\emph{Introduction:---} Floquet symmetry-protected topological (FSPT) phases  are emergent condensed matter phenomena~\cite{kitagawa2010topological,jiang2011majorana,rudner2013anomalous,else2016classification,potter2016classification,von2016phase1,von2016phase2,potirniche2017floquet,roy2016abelian} that extend the concept of symmetry-protected topological (SPT) order to periodically driven systems~\cite{chen2011classification,chen2011complete,chen2012symmetry,chen2013symmetry}. 
A key aspect of one dimensional SPTs is having ground states with protected degeneracies in their entanglement spectrum \cite{pollmann2010entanglement,cornfeld2019entanglement,azses2020identification,azses2020symmetry,de2020inaccessible}. For unitary symmetries, these degeneracies can be detected with symmetry-resolved entanglement (SRE) measures~\cite{goldstein2018symmetry,xavier2018equipartition,cornfeld2018imbalance,bonsignori2019symmetry,feldman2019dynamics,horvath2020symmetry,fraenkel2020symmetry,neven2021symmetry, fraenkel2021entanglement}, and allow to use SPTs as universal computational resources \cite{else2012symmetry,stephen2017computational,raussendorf2001one}. Whether and how these properties show up for FSPT order is a question that we address in this paper. In periodically driven systems, the key object that admits topological features is the unitary Floquet operator describing the time evolution for one cycle, $F = U(T,0)$, where $T$ is the time period. Its eigenvalues, $\lambda_i = e^{-i T \varepsilon_i}$, which define the quasienergies $\omega_i = T \varepsilon_i \ {\rm mod} \ 2\pi$, have topological characteristics such as protected $0-$ and/or $\pi-$ edge modes~\cite{jiang2011majorana}. 

\begin{figure}[b]
	\includegraphics[width=\linewidth]{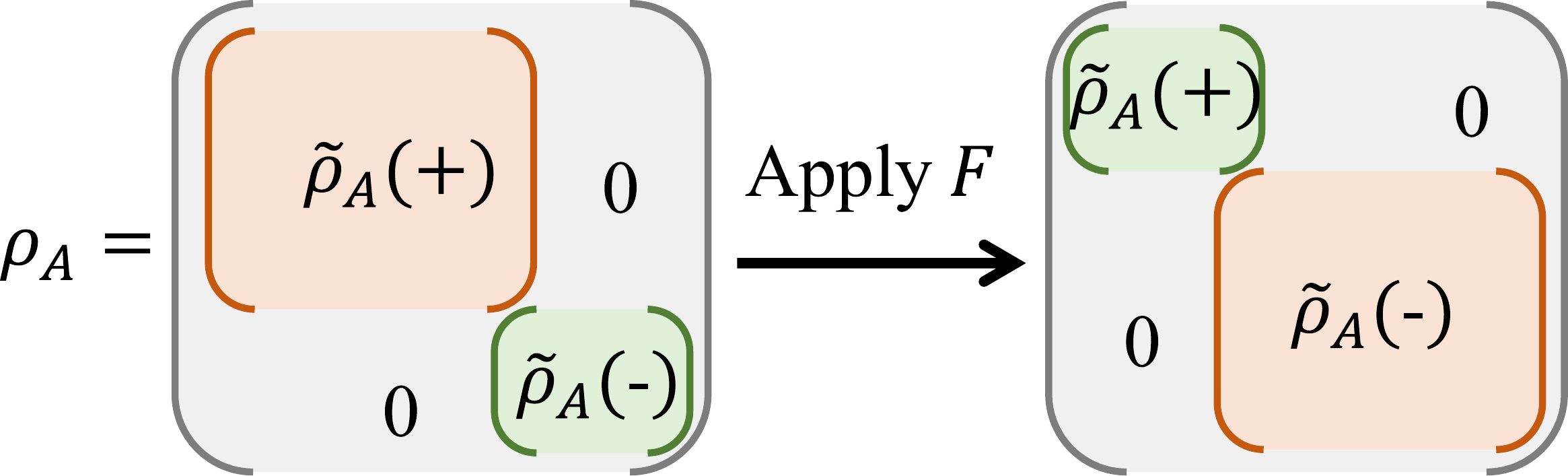}
	\caption{FSPT order is characterized by cyclic switching of symmetry blocks of the reduced  density matrix upon applying the Floquet operator $F=U(T,0)$.
	}
	\label{fig:f_sre_diag}
\end{figure}

In contrast to static SPTs, the eigenstates of $F$ are not necessarily entangled, even in nontrivial FSPTs. Instead, entanglement in FSPTs is hidden in the time evolution within a period, which is often referred to as {\it micromotion} and is generically characterized by quantized charge pumping ~\cite{kumar2017string,potter2016classification}. Here, we study and experimentally observe this phenomenon by focusing on the dynamics of the SRE, derived from the block diagonal structure of the reduced density matrix $\rho_A={\rm{Tr}}_B\rho$ \cite{azses2020symmetry}. In the static case, the SRE structure can be used to identify SPT phases via degeneracies between the symmetry blocks~\cite{azses2020identification}. Our key observation here is that nontrivial FSPT order is reflected by an exact  cycling of the symmetry blocks upon Floquet  evolution, as  illustrated in Fig.~\ref{fig:f_sre_diag}. As an experimentally detectable~\cite{azses2020identification,vitale2021symmetry} consequence, the first moment of the SRE, defined as the subsystem charge, displays cyclic switching.
This is demonstrated for a $\intz_2$ FSPT phase on a noisy intermediate-scale quantum (NISQ) computer. We quantify the parity switching as an order parameter, and observe its dynamics across a Floquet topological phase transition. We propose a generalization of measurement-based quantum computation (MBQC) to the FSPT case. Lastly, static SPT order can also coexist with nontrivial Floquet order, in which case the protected entanglement is associated with degeneracies between cyclically switching symmetry blocks.

\emph{Cohomological classification:---}
Before discussing our main result, we put it in the mathematical context of the classification of 1D bosonic SPTs. A 1D SPT phase protected by the symmetry group $G$, is characterized by a ground state accompanied by a symmetry operator $U(g)$ representing the group $G$. While $U(g)$ acts on the full system as a conventional representation, it acts near the edges~\cite{else2012symmetry,perez2008string} via a projective representation that classifies the different SPT phases into $\calh^2[G,U(1)]$ classes~\cite{chen2013symmetry}.

One-dimensional bosonic FSPTs are characterized by an additional discrete symmetry, namely translations in time by integer multiples of the period, or equivalently, discrete powers of $F$. Because this operator commutes with the static symmetry $G$, the total system is characterized by a $G\times \intz$ symmetry. As a result, there are $\calh^2[G\times \intz,U(1)]$ bosonic FSPT phases~\cite{else2016classification, potter2016classification,tantivasadakarn2021symmetric}. For finite Abelian groups $G$ we find that~\cite{SM} (see, also, references \cite{pontryagin1934theory,kampen1935locally,berkovich1998characters} therein)
\be
\label{eq:classification}
\calh^{2}[G \times \intz,U(1)]= \calh^{2}[G,U(1)] \times G.
\ee
One can understand the two factors in Eq.~(\ref{eq:classification}) as a bulk SPT order classified by $\calh^{2}[G,U(1)]$, which results in degeneracies between the symmetry blocks of $\rho_A$, and additional $|G|$ phases that characterize the possible cyclic permutations of the SRE, after applying $F$. Importantly, even symmetry groups whose cohomology group is trivial and cannot support static SPT phases can protect nontrivial Floquet topology.

\emph{SRE switching:---} For a system characterized by a unitary symmetry $G$ with a conserved charge $\calq_{tot}$, the density matrix of the reduced system $A$ has a decomposition~\cite{laflorencie2014spin} $\rho_A = \oplus_\calq \tilde{\rho}_A(\calq)$ associated with subsystem charge $\calq \equiv \calq_A$. We define the $n$'th \Renyi SRE as $S_n (\calq)= {\rm{Tr}}[\Pi_{\calq} \rho_A^n ]$, where $\Pi_{\calq}$ projects subsystem $A$ to charge sector $\calq$. For example, for the symmetry group $G=\intz_N$, the charge $\calq$ is an integer, modulo $N$. This group has a trivial cohomology group, $\calh^{2}[\intz_N,U(1)]$ and, hence, cannot support static SPTs. According to Eq.~\ref{eq:classification}, we have exactly $|\intz_N|=N$ distinct FSPTs phases. Each phase is labeled by an integer  $c=0,1,...,N-1$, which represents the pumped charge of the FSPT phase. Let us focus on eigenstates of the Floquet operator in the bulk, but not necessarily on the edges. After one cycle $\rho \to \rho' = F \rho F^\dagger$ and, as we now demonstrate,
\be
\label{eq:cyc_sre}
S_n(\calq) \to S'_n(\calq) = \Tr[ {\rho_A'}^n  \Pi_\mathcal{\calq}   ] = S_n(\calq'),
\ee
where $\calq'=\calq+c$. Equation~\ref{eq:cyc_sre} tells us that the symmetry sector $\calq$ goes to  $\calq'$  upon acting with $F$. Being valid for any $n$, this relation implies the cycling of the entire spectrum of each symmetry block $\tilde{\rho}_A(\calq)$, as schematically shown  in Fig.~\ref{fig:f_sre_diag}. 
\begin{figure*}[t]
	\includegraphics[width=\linewidth]{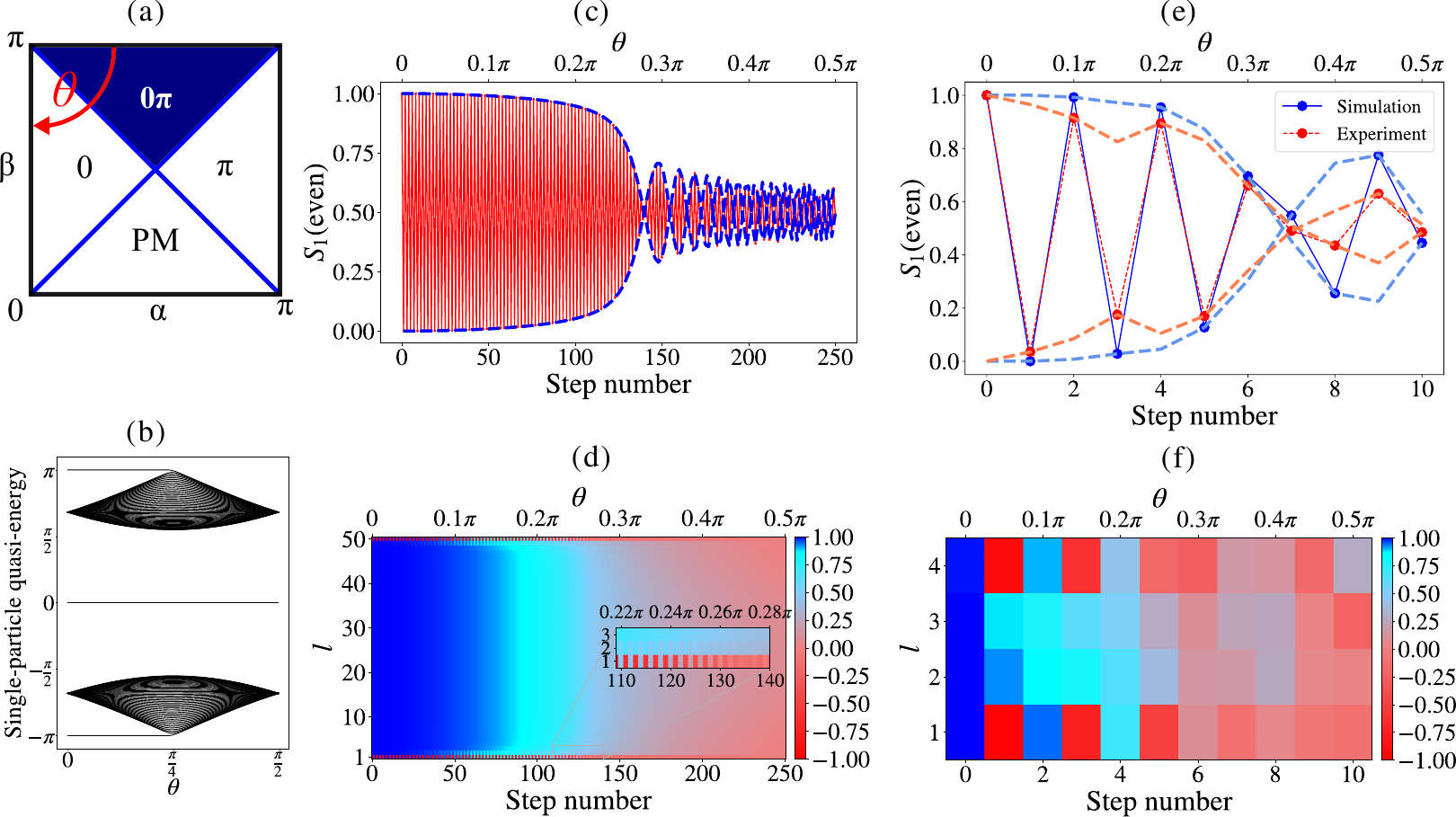}
	\caption{(a) Phase diagram of the model (\ref{eq:UIsing}) reproduced from \cite{khemani2016phase}. The phases PM, $0$, $\pi$, and $0\pi$ are, respectively, the paramagnet,  ferromagnet,  time crystal, and  topological phase.  Dark blue represents the topological phase exhibiting SRE switching. The red curve indicates the adiabatic path which we transverse in $N_{\rm steps}$ steps. (b) Single particle quasi-energy spectrum along the  path in (a) exhibiting a $\pi-$mode for $\theta \le \pi/4$. (c,d) Exact calculation of subsystem parity probability $S_1({\rm{even}})$ and $\langle X_l \rangle$ for $L=2L_A=50$ and $N_{\rm steps}=250$. $S_1({\rm{even}})$ in (c) displays parity switching in the topological ($0\pi$) phase up to the phase transition at step $\sim N_{\rm steps}/2$ followed by a beating structure in the $(0)$ ferromagnetic phase. $\langle X_l \rangle$ in (d) displays edge state parity switching (see inset) in the topological phase. (e,f) Same quantities as in (c,d) measured on a NISQ computer for $L=2L_A=4$ and $N_{\rm steps}=10$.}
	\label{fig:phasediagram}
\end{figure*}

We now prove this result using the framework of Ref.~\cite{von2016phase1}. Assuming that the FSPT phase with symmetry $G$ is  integrable or many-body localized (MBL)  (see Refs.~\cite{abanin2015theory,abanin2019colloquium,bahri2015localization}), we have an edge-bulk decomposition of the one-period time evolution operator
\be
\label{eq:f_decomp}
F = v_Lv_R e^{-if}, 
\ee
where $v_L$ ($v_R$) is a unitary operator localized at the left (right) part of the system and $f$ is a functional of the symmetric (w.r.t. $G$) constants of motion associated with the MBL \cite{von2016phase1}. Additionally, we have the identity 
\bea
\label{eq:sym_1d_rep}
U(g)v_LU^{\dagger}(g)v_L^\dagger = \kappa_c(g),  \\
U(g)v_RU^{\dagger}(g)v_R^\dagger = \kappa_c^{-1}(g), \nonumber
\eea
for unitary symmetries represented by $U(g)$, where $g \in G$ and $\kappa_c(g)$ is the $1D$ representation of the FSPT matching the group element $c$ such that $U(c)\ket{g} =\kappa_c(g)\ket{g}$ \cite{von2016phase1}. As the state is an eigenstate of $F$ in the bulk, it is an eigenstate of $e^{-if}$. If we consider Floquet  phases that do not break any symmetry (including time translations) the state is an eigenstate of $e^{-if}$ and this phase factor drops out of the evolved state $\rho' = F \rho F^\dagger = v_L v_R \rho v_R^\dagger v_L^\dagger$ \cite{von2016phase1}. 

We now calculate the first moment $S_1(\calq) = \Tr[\Pi_{\calq} \rho_A]$ of the SRE, which is the probability that the reduced system has charge $\calq$. For Abelian unitary finite symmetries,  the projectors can be written in terms of the group characters $\chi_{\calq}(g)$ \cite{yen2019exact, de2020inaccessible},
\be
\label{charac}
\Pi_\mathcal{\calq}=\frac{1}{|G|} \sum_{g \in G} {\chi_\calq(g)} U_A(g), 
\ee
where $U_A(g)$ is the symmetry acting on the reduced system $A$. If subsystem $A$ includes the left but not the right edge, we have that $\rho_A \to \rho_A' = v_L \rho_A v_L^\dagger$ and $\Pi_\mathcal{\calq}$ commutes with $v_R$. Thus, after one cycle $S_1(\calq) \to S_1'(\calq)  = \Tr[v_L^\dagger \Pi_\calq v_L \rho_A ]  =\frac{1}{|G|} \sum_{g \in G} {\chi_\calq(g)} \Tr[ v_L^\dagger U_A(g) v_L   \rho_A ]$. Using the identity $v_L^\dagger U_A(g) v_L = \kappa_c(g)U_A(g)$, which derives from Eq.~\ref{eq:sym_1d_rep} by partial tracing, we obtain
\be
\label{projTrans}
v_L^\dagger \Pi_{\calq} v_L= \Pi_{\calq + c}.
\ee 
Then Eq.~(\ref{eq:cyc_sre}) follows for the case $n=1$. Generalizing this result to any $n$ is straightforward: After one cycle, we have that $\rho_A^n \to (v_L\rho_A v_L^\dagger)^n = v_L\rho_A^n v_L^\dagger$ as $v_L$ is unitary and satisfies $v_L^\dagger v_L = I$. As a result, the $n$'th \Renyi SRE evolves as $S_n(\calq) = \Tr[\Pi_\calq \rho_A^n ]\to S'_n(\calq) = \Tr[v_L^\dagger \Pi_\calq v_L \rho_A^n ]=\frac{1}{|G|} \sum_{g \in G} {\chi_\calq(g)} \Tr[ v_L^\dagger U_A(g) v_L  \rho_A^n ] $. Together with the assumptions and derivations above, this proves Eq.~\ref{eq:cyc_sre} for any $n$.

\emph{Parity switching through a phase transition:---} The subsystem charge can serve as a measurable order parameter of FSPT phases that can be observed even on small noisy quantum computers. We now study its evolution across a phase transition.

To showcase the use of this order parameter, we consider the kicked Ising model $F(\alpha,\beta)=U_{ZZ}(\beta)U_X(\alpha)$ where
\bea
\label{eq:UIsing}
U_{X}(\alpha) &= e^{i\frac{\alpha}{2} \sum_{l=1}^{L} X_l }, \\
U_{ZZ}(\beta) &= e^{i\frac{\beta}{2} \sum_{l=1}^{L-1} Z_lZ_{l+1}}, \nonumber
\eea
and $X_l,Z_l$ are Pauli matrices acting on the $l$'th site of a  chain with open boundary conditions. This model has a $G=\intz_2 $ symmetry represented by the parity operator $P = \prod_{l=1}^{L} X_l$. Its phase diagram~\cite{khemani2016phase} is displayed in Fig.~\ref{fig:phasediagram}(a). The four phases are labeled by the number of single Majorana fermion excitations at quasi-energies $0$ and $\pi$, see Fig.~\ref{fig:phasediagram}(b). The phases with an odd number of Majoranas spontaneously break the Ising symmetry, and correspond to a ferromagnet (0 phase) and to a time crystal ($\pi$ phase)~\cite{else2016floquet,verresen2017one,friedman2017phases,khemani2019brief}. The latter phase was realized using trapped ions~\cite{zhang2017observation} and superconducting circuits \cite{khemani2019brief,ippoliti2020many,frey2021simulating,mi2021observation}. The Floquet topological phase $0\pi$ was realized in Ref.~\cite{potirniche2017floquet} using cold atoms.

Here we focus on the transition from the FSPT $(0\pi)$ phase to the ferromagnetic $(0)$ phase. The former phase has two interconnected properties: protected edge states at quasi-energy $\pi$ and SRE swapping. The probability of the even subsystem parity $S_1({\rm{even}}) = \frac{1+ \langle P_A \rangle}{2}$ (with $S_1({\rm{odd}}) = 1-S_1({\rm{even}})$) is obtained by evaluating the subsystem parity $P_A = \prod_{i \in A} X_i$ where $A$ includes $L_A$ sites from the left boundary. In the topological phase, $S_1({\rm even})$ and $S_1({\rm odd})$ are expected to swap their values at each time step. This behavior is trivially seen, for example, at the sweet spot $\beta=\pi$ and $\alpha=0$ where $U_{ZZ}(\pi) =(-i)^{L-1} Z_1 Z_L$ and $U_X=1$. Then the evolution over one period $F=U_{ZZ} U_X$ simply flips the two edges from $+$ to $-$ and  $S_1({\rm even})$ and $S_1({\rm odd})$ alternate between 0 and 1.

To see that this property persists in the entire topological phase, but disappears in the ferromagnetic phase, we adiabatically change the parameters of $F$. Specifically, we follow the path $\alpha=r_0 \cos \theta$ and $\beta=\pi-r_0 \sin \theta$, shown as a red curve in Fig.~\ref{fig:phasediagram}(a), in $N_{\rm steps}$ equal steps, such that the topological phase transition is crossed at $\theta=\pi/4$, i.e. at the $N_{\rm steps}/2$ step. We set $r_0=1$ throughout, and initialize the system in the state $|+\rangle = \otimes_l|+\rangle_l$.

The switching of $S_1({\rm{even}})$ is shown along an adiabatic protocol with $N_{\rm steps}=250$ in Fig.~\ref{fig:phasediagram}(c) for a system with $L=50$ qubits and $L_A=L/2$~\cite{SM} (see, also, references \cite{kitaev2001unpaired,aguado2017majorana,terhal2002classical,wimmer2012algorithm} therein). Fig.~\ref{fig:phasediagram}(d) gives a real space picture of $\langle X_l \rangle$, showing that for small $\theta$, the edge spins are responsible for the switching. As approaching $\theta=\pi/4$, the extent of the switching zone increases, and eventually the two edges merge at the Floquet topological phase transition. 

For $\theta>\pi/4$, $S_1({\rm{even}})$ gives rise to a beating structure, persisting into the ferromagnetic phase. The frequency of this beating is determined by the difference between two quasi energies of the instantaneous Floquet operator $F(\theta)$. Specifically, our initial state $ |+ \rangle $ can be combined with the state $Z_1 Z_L |+ \rangle $ to form a pair of Floquet eigenstates of $F(\theta=0)$, denoted by $|1\rangle$ and $|2\rangle$, with quasienergies $\omega_{1}$ and $\omega_{2} = \omega_{1} + \pi$. As we vary $\theta$ adiabatically, the quasienergies follow their instantaneous values $\omega_{1,2}(\theta)$, see Fig.~\ref{fig:phasediagram}(b) for the single-particle quasienergies. In the thermodynamic limit ($L\to\infty$), $\omega_2(\theta)-\omega_1(\theta)$ is pinned to $\pi$ in the topological phase, giving rise to a periodic switching of $S_1({\rm even})$. For finite systems, there are small deviations, which become significant near the phase transition. Moving  into the ferromagnetic phase $\omega_2- \omega_1$ deviates from $\pi$ and eventually becomes the energy difference of a domain wall in the ferromagnetic order ~\cite{SM}. This picture allows us to show that the beating amplitude vanishes in the thermodynamic limit as $1/\sqrt{L}$~\cite{SM}, implying that the parity switching is a FSPT order parameter. The parity switching is observable on a quantum computer as shown in Fig.~\ref{fig:phasediagram}(e,f) for $L=4$ and $L_A=L/2$ and $N_{\rm steps}=10$ steps \cite{SM}.

\begin{figure}[t]
	\includegraphics{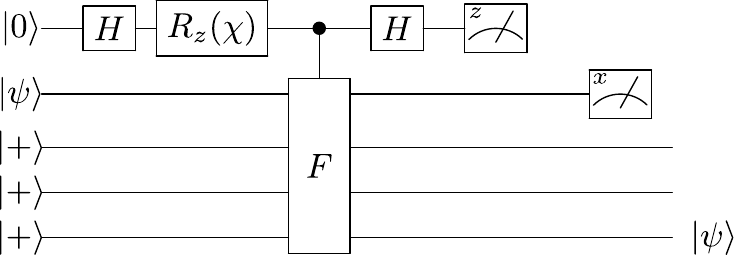}
	\caption{Teleportation protocol as a probe of FSPT order. A state $| \psi \rangle$ is prepared on the edge. A controlled-$F$ operator followed by a measurement of an ancilla qubit  entangles the edge states, allowing for teleportation upon measurement of the first qubit, iff $F$ is topological. }
	\label{fig:circuit_mbqc}
\end{figure}

\emph{Teleportation  through a FSPT phase:---} The existence of a Floquet operator that switches the symmetry blocks of the reduced density matrix leads to computational power. In the case of SPTs, computation power relies on their entanglement~\cite{raussendorf2001one,raussendorf2017symmetry}; By measuring the state in  specific bases, the correlations that compose the entanglement cause the input information to flow. Here, we discuss how FSPT order, with generically unentangled eigenstates, allows to teleport a quantum state.

We, again, focus on the simplest case of $G=\intz_2$. Our basic procedure is composed of three steps: (i) Start from a Floquet eigenstate and encode a ``qubit" state $\ket{\psi} = \sum_{g} \alpha_g | g \rangle$ at the left edge in the $G$-symmetry eigenbasis $|g \rangle = |\pm \rangle$. Encode the identity group element  state $|+\rangle$ in the right edge qubit. (ii) Use an ancilla qubit to apply $I+ e^{i \chi} F$  with a nonuniversal phase factor $e^{i \chi} = e^{i f} $ of the Floquet eigenstate. (iii) Lastly, measure the left edge qubit in the $G$-symmetry basis. If the FSPT state is nontrivial, then the left edge qubit $|\psi \rangle$ is teleported to the right. This property follows generally for the $\intz_N$ case ~\cite{SM} directly from the algebra in Eqs.~(\ref{eq:f_decomp}) and (\ref{eq:sym_1d_rep}) which defines the action of $F$ and of the symmetry projectors in a given FSPT phase.

This protocol is depicted in Fig.~\ref{fig:circuit_mbqc} for the point $\theta=0$ in the topological phase where the edge states are localized on the edge qubits. In this case the initial quantum state is  $|\psi \rangle_1 \otimes_{i=2}^L |+\rangle_i$, where $|\psi \rangle_1=\alpha_+ |+\rangle_1+ \alpha_- |-\rangle_1$. We have $F=Z_1 Z_L$ with some additional phase $\chi=\frac{\alpha}{2}(L-2)$. It is easy to see that after applying $I+e^{i\chi}Z_1 Z_L$ and then projecting the left qubit, say to $|+ \rangle_1$, the final state is $\otimes_{i=1}^{L-1} |+\rangle_i  |\psi \rangle_L$. In Fig.~\ref{fig:circuit_mbqc} we implement the operator $I+e^{i\chi}F$ using a simple quantum circuit. Adding an ancillary qubit $\ket{0}$ and then acting with Hadamard gate $H$ on it we get the state $\frac{1}{\sqrt{2}}(\ket{0}+\ket{1})$. Then, we add the phase by rotating around the $Z$-axis with $R_z(\chi) = e^{-i \frac{\chi}{2} Z}$, which takes the ancillary qubit to the state $\frac{1}{\sqrt{2}}(\ket{0}+e^{i\chi}\ket{1})$. Next, by applying a controlled-$F$ gate, we only apply $F$ on the initial state for the $\ket{1}$ ancilla state, thus, the whole circuit is in the state $\frac{1}{\sqrt{2}}(\ket{0} \otimes \ket{\psi}+\ket{1}\otimes e^{i\chi}F\ket{\psi})$. Lastly, applying $H$ again on the ancillary qubit, we have $\frac{1}{2}(\ket{0} \otimes [I + e^{i\chi}F]\ket{\psi} + \ket{1}\otimes [I - e^{i\chi}F]\ket{\psi} )$. It is now clear that in the case of measuring the ancilla in the $\ket{0}$ state we implement the operator $I+e^{i\chi}F$ on the initial state. As in MBQC, one can correct for the possible measurement outcomes, and also perform general rotations by measuring in a rotated basis. In order to probe any other point in the FSPT phase, one can apply our adiabatic protocol on the initial state, transforming $\theta =0 \to \theta_1 < \pi/4$, then apply $I+F(\theta_1)e^{i\chi(\theta_1)}$, and  adiabatically evolve back to $\theta=0$ where the final measurement is done on the first qubit.

\emph{Entanglement switching:---} So far, we considered only unentangled FSPT states with a trivial static cohomology group $\calh^2[G,U(1)]$. For non-trivial groups, the Floquet eigenstates contain protected entanglement linked with degeneracies of $\rho_A$ between different symmetry sectors \cite{azses2020symmetry}. Consider for example $G=\intz_N \times \intz_N$, where static SPTs are classified by $m \in \calh^{2}[G,U(1)] = \intz_N$ \cite{chen2013symmetry} $(m=0,\dots N-1)$. In the presence of a Floquet drive classified by an element $c=(c_1,c_2) \in  G$ as in Eq.~(\ref{eq:classification}), the block $\calq$ turns into $\calq+c$. The degenerate blocks $\calq = (q_1,q_2)$ are grouped into families whose representative is $q'=(q'_1,q'_2)$, defined $ {\rm{mod~gcd}} (N,m)$.  The entanglement switching of these families is described by $q'_i \to q'_i + c_i~ {\rm{mod~gcd}} (N,m) $ ($i=1,2$)~\cite{SM}. We discuss examples of this general formula. If $m$ and $N$ have no common divisors greater than $1$, all the symmetry blocks are degenerate and there is no nontrivial  switching under the action of $F$. In contrast, in SPT static phases where $m > 1$ divides $N$ (for example $m=2$, $N=4$) the degeneracy is only partial~\cite{azses2020symmetry}, leading to entanglement, and the  degenerate blocks switch from one family to another~\cite{SM}.

\emph{Summary:---} The global topological properties of 1D FSPT phases cannot be revealed by any local measurements in the bulk. Here, we used symmetry resolution measurements to observe the FSPT order, both on a small system realized by a NISQ computer, and analytically on large systems. The latter allowed us to describe a Floquet phase transition into a topologically trivial phase. The topological edge excitations of the Floquet phase, adiabatically evolve to domain wall excitations of a ferromagnet. Conversely, this property allows one to prepare adiabatically topological excitations, starting from local excitations. Finally, we demonstrated the ability of FSPT order to teleport a quantum state. All these topological properties, similar to the cohomological classification, hold for periodically driven interacting bosons in general and are not limited to our showcase kicked Ising model which admits a free fermion description.

\emph{Acknowledgments:---}  ES acknowledges support from  ARO (W911NF-20-1-0013), European Research Council (ERC) under the European Unions Horizon 2020 research and innovation programme under grant agreement No. 951541, and the US-Israel Binational Science Foundation (Grant No. 2016255). This work was supported by the Israel Science Foundation, grants number 151/19 and 154/19, and the AWS Cloud Credit for Research Program. We thank discussions with Alon Ron. We acknowledge the use of QuTiP Python library for some of the numerical calculations \cite{JOHANSSON20121760,JOHANSSON20131234}

\section*{Supplemental Materials}
\section{FSPT characterization and cohomology classification}
\label{app:fspt_coho}
In this appendix we derive Eq.~(1) of the main text. In one dimension, we have simple formulas for FSPTs of bosons for finite Abelian groups. We start by reviewing known results. Let $G$ be a finite Abelian group, then %it is well known that 
the Pontryagin duality indicates that $\calh^1[G,U(1)] \cong G$ \cite{pontryagin1934theory,kampen1935locally}. Additionally, it is known that $\calh^i[G,U(1)] = \calh^{i+1}[G,\intz]$, $\calh^1[\intz,\intz]=\calh^0[\intz,\intz]=\intz$ and $\calh^j[\intz,\intz]=0$ for $j\neq0,1$ \cite{chen2013symmetry,site2021free}. Moreover, the cohomology of $\calh^d[A \times B,\intz]$ can be split in terms of the cohomology of $\calh^i[A,\intz]$ and $\calh^i[B,\intz]$ with $i\leq d$ according to Knuth formula, see Eq.~(J55) of Ref.~\cite{chen2013symmetry}. In the case of $A=G$ and $B=\intz$, we have
\begin{eqnarray*}
	\calh^d[G \times \intz,U(1)]&=& \\
	\calh^{d+1}[G \times \intz,\intz] &=& \calh^{d+1}[G,\intz] \otimes_{\intz}    \calh^0[\intz,\intz]  \times\\
	& & \calh^{d}[G,\intz] \otimes_{\intz}    \calh^1[\intz,\intz]  \times\\
	& & \mathrm{Tor}_1^\intz[\calh^{d+1}(G,\intz), \calh^0(\intz,\intz)] \times\\
	& & \mathrm{Tor}_1^\intz[\calh^{d}(G,\intz), \calh^1(\intz,\intz)], \\
\end{eqnarray*}
where we have used the fact that $\calh^i[\intz,\intz]  \otimes_{\intz} M = \intz_1 \otimes_{\intz} M = \intz_1$ for $i\geq 2$ (where $M$ is the corresponding cohomology group) according to Eq.~(J51) of Ref.~\cite{chen2013symmetry}, and we have also used similar relations for $ \mathrm{Tor}_1^\intz[M,\intz_1] = \intz_1$ and thus we have omitted terms containing $\calh^i[\intz,\intz]$ with $i\geq 2$. The remaining $\mathrm{Tor}_1^\intz$ terms are also $0$ as a result of $\mathrm{Tor}_1^\intz[M,\intz]=0$. Applying all the results from the start of the paragraph with Eq.~(J51) from Ref.~\cite{chen2013symmetry} we end up with $\calh^{d+1}(G \times \intz,\intz)= \calh^{d+1}(G,\intz) \times  \calh^{d}(G,\intz)$. Therefore, we conclude that $\calh^{d}[G \times \intz,U(1)]= \calh^{d}[G,U(1)] \times  \calh^{d-1}[G,U(1)]$. For the case of one dimension, we have that
\be
\calh^{2}[G \times \intz,U(1)]= \calh^{2}[G,U(1)] \times G,
\ee
where we have used that $\calh^{1}(G,U(1)) \cong G$ for finite Abelian group $G$.

\section{Mapping to free fermion model}
\label{app:ferm_map}
In this appendix we detail the free fermion methods used to compute the spectrum in Fig.~2(b) and parity switching in Fig.~2(c) of the main text. Our $\intz_2$ bosonic model can be exactly mapped to free fermion model by applying the Jordan-Wigner (JW) transformation. The resulting fermionic chain, often referred to as the Kitaev chain \cite{kitaev2001unpaired,aguado2017majorana} can be solved by going into the single-particle picture, whose computational dimension is reduced exponentially.

The JW transformation maps the stroboscopic evolution to free-fermion single-particle stroboscopic evolution. The evolution $e^{-iH_X} = e^{-i \frac{\alpha}{2} \sum_i X_i}$ is mapped into an on-site term $e^{-i \alpha \sum_i (\frac{1}{2}-f_i^\dagger f_i^{\phantom{\dagger}})}$ and the interaction $e^{-iH_{ZZ}} = e^{-i \frac{\beta}{2} \sum_i Z_i Z_{i+1}}$ is mapped into an hopping term $e^{-i \frac{\beta}{2} \sum_{j} f_j^{\dagger} f_{j+1}^{\phantom{\dagger}}+f_{j+1}^{\phantom{\dagger}} f_{j}^{\phantom{\dagger}} + {\rm c.c.}}$, where $f_i^\dagger,f_i^{\phantom{\dagger}}$ are the creation and annihilation fermionic operators. The resulting fermionic Hamiltonian can be easily solved by introducing the Majorana fermion operators. Defining $f_i = \frac{\gamma_{2i}+i\gamma_{2i+1}}{2}$ we obtain a new Hamiltonian $H_M =\sum_{i,j} \gamma_i A \gamma_j  \equiv -i \sum_{i,j} \gamma_i \frac{h}{4} \gamma_j$, where $A$ is an anti-symmetric matrix (as the Hamiltonian is hermitian) and $h$ is the single-particle Hamiltonian (see Eq.~(57) in Ref.~\cite{aguado2017majorana} for the exact Majorana form). The operators mappings are $X_i \to -i\gamma_{2i}\gamma_{2i+1}, Z_iZ_{i+1} \to -i\gamma_{2i+1}\gamma_{2(i+1)}$ obtained by direct computation. By focusing on the single-particle Hamiltonian $h$ we reduce the dimension of the Hamiltonian to $2L \times 2L$, obtaining an efficient way to compute the time evolution. Combining two such single-particle Hamiltonians $h_1,h_2$ into an effective single-particle Hamiltonian $h_3$ is given by $e^{-i h_3} = e^{-ih_2} e^{-ih_1}$, which can be proved by the BCH formula and the commutation relations of the Majorana operators. This combined Hamiltonian satisfy the relation $e^{-i H_m^3} = e^{-i H_m^2}e^{-i H_m^1}$, where $H_m^i$ is the Majorana Hamiltonian generated by the corresponding $h_i$ single-particle Hamiltonian. Therefore, the stroboscopic evolution in the Majorana picture can be solved effectively for many periods by combining the single-particle Hamiltonians iteratively.

\subsection*{Majorana exact solution}
We proceed to calculate the average subsystem parity $\langle X_1\dots X_{L_A} \rangle$ using the explicit Majorana formalism of the non-interacting fermions in $3$ steps: (i) The average parity transforms in the Majorana picture is given by the average of $(-i)^{L_A} \langle \gamma_{0}\gamma_{1}...\gamma_{2L_A-1} \rangle = (-i)^{L_A} {\rm Pf}(M)$, where $M_{ij}$ is the anti-symmetric $2L_A \times 2L_A$ matrix $\langle \gamma_i \gamma_j \rangle$ with zero diagonal for $i,j=0,\dots,2L_A-1$ and ${\rm Pf}(M)$ denotes the Pfaffian, see Ref.~\cite{terhal2002classical} (For the calculation of the Pfaffian, we have used the Python library pfapack, which implemented the fast algorithm in Ref.~~\cite{wimmer2012algorithm} for calculating Pfaffians). (ii) We continue from here by calculating the $\langle \gamma_i \gamma_j \rangle$ on the time-evolved state by observing the form of $U^\dagger \gamma_i U$. As the total effective Hamiltonian is quadratic, we have a simple form for the Hadamard formula
\be
\label{eq:hadamard}
e^XYe^{-X} = Y + [X,Y] + \frac{1}{2!} [X,[X,Y]] + \dots,
\ee
where the action $[X,O]$ on operator $O$ is known as the adjoint $\text{Ad}_{X} Y$. (iii) Putting it all together, we obtain the parity for very large $L$'s as this method scales polynomially with $L$.

In our case, we have that $X = \sum_{i,j} A_{ij} \gamma_i \gamma_j, Y = \gamma_k$. Therefore, we only need to compute $[\gamma_i\gamma_j,\gamma_k]$. The result is given from simple commutation identities $[AB,C] = A[B,C] + [A,C]B$. Hence, $[\gamma_i\gamma_j,\gamma_k] = 2[\gamma_i (\delta_{jk}-\gamma_k\gamma_j) + (\delta_{ik}- \gamma_k\gamma_i) \gamma_j] = 2[\gamma_i \delta_{jk}-\gamma_j \delta_{ik}]$, where we used $\{\gamma_i,\gamma_j\} = 2\delta_{ij}$ and $[\gamma_i,\gamma_j] = \gamma_i\gamma_j - \gamma_j\gamma_i = 2(\delta_{ij} -\gamma_j\gamma_i)$. The result is that $[X,Y] = \sum_{ij} A_{ij} [\gamma_i \gamma_j,\gamma_k] =  \sum_{ij} 2A_{ij} [\gamma_i \delta_{jk}-\gamma_j \delta_{ik}] = 2 (\sum_{i} A_{ik} \gamma_i - \sum_{j} A_{kj} \gamma_j) = 2\sum_{i} (A_{ik}- A_{ki}) \gamma_i$.

We continue by calculating $[X,[X,Y]]$. Substituting the last result we have that $[X,[X,Y]] =  \sum_{i,j,l} A_{ij} 2(A_{lk}- A_{kl}) [\gamma_i\gamma_j,\gamma_l] = \sum_{i,j,l} 2A_{ij}(A_{lk}- A_{kl}) 2[\gamma_i \delta_{jl}-\gamma_j \delta_{il}] = \sum_{i,j} 4A_{ij}[(A_{jk}- A_{kj}) \gamma_i -\gamma_j (A_{ik}- A_{ki})] = 4 \sum_{i,j} [A_{ij}- A_{ji}][A_{jk}- A_{kj}]\gamma_i $. More generally, assuming $Y = \sum_l B_l \gamma_l$, we see that applying ${\rm Ad}_X$ (i.e., $Y \rightarrow [X,Y]$) on it, we have that $B \rightarrow 2(A-A^T)B$ (it transforms a vector to vector) after such adjoint action. We derive it explicitly, $[X,Y] = \sum_{ijl} A_{ij} B_l [\gamma_i \gamma_j,\gamma_l] =  2\sum_{ijl} A_{ij} B_l [\gamma_i \delta_{jl}-\gamma_j \delta_{il}] = 2\sum_{ij} A_{ij} B_j\gamma_i - 2\sum_{ij} A_{ij} B_i\gamma_j = 2\sum_{i} [\sum_j (A_{ij}-A_{ji}) B_j ]\gamma_i $. Therefore, we have that $B_i \rightarrow 2(A_{ij}-A_{ji}) B_j = 2((A-A^T)B)_i$. Thus, the adjoint action corresponds to a transformation $B \rightarrow  2(A-A^T)B$.

Using the formula from the last paragraph we sum the series given by Eq.~\ref{eq:hadamard}. We have that $e^X Y e^{-X} = \sum_n \frac{2^n}{n!}[(A-A^T)^n B]^i \gamma_i$ (we use Einstein summation convention here and throughout the section) for $X = \sum_{ij} A_{ij} \gamma_i \gamma_j$, $Y = \sum_i B_i \gamma_i$. For anti-symmetric $A=-A^T$ we can simplify the expression $e^X Y e^{-X} = \sum_n \frac{4^n}{n!} [A^nB]^i\gamma_i = [e^{4A}B]^i \gamma_i$. Hence, we are able to calculate exactly $\langle \gamma_i \gamma_j \rangle$ using the above derived formula, which uses only $2L \times 2L$ matrices, and thus exponentially faster.

To complete the calculation we expand $\langle \psi |\gamma_i \gamma_j | \psi \rangle$ with linear combinations of $\langle 0 |\gamma_{i'} \gamma_{j'} | 0 \rangle$ and get a simple formula for the vacuum expectation values. First, using the results of the last paragraph we can simply expand $\langle \psi |\gamma_i \gamma_j | \psi \rangle = \sum_{i',j'}[e^{4A}\delta_{k,i}]^{i'} [e^{4A}\delta_{k,j}]^{j'} \langle 0| \gamma_{i'} \gamma_{j'}|0\rangle $. Thus, we are left to calculating the vacuum expectation values $\langle 0| \gamma_{i'} \gamma_{j'}|0\rangle $. In the fermionic language, the state $\ket{0}$ is mapped into the even parity fermionic state over pairs, which is the eigenstate of $-i\gamma_{2i}\gamma_{2i+1}$ with eigenvalue $+1$. Applying $\gamma_k$ on such even parity state will break its parity for the respective pair, but when accompanied with its pair the even parity is restored, thus, $\langle 0| \gamma_{i'} \gamma_{j'}|0\rangle \propto \delta_{i',j'\pm1}$ for $i' \neq j'$, where the sign depends on how the pairs are clustered. We can also show this analytically by explicitly casting the Majorana operators to fermionic ones $\gamma_{2i} = f_i + f_i^{\dagger}, \gamma_{2i+1} = -i (f_i -f_i^{\dagger}) $. Let us consider all cases: 1. $\langle 0 | \gamma_i \gamma_i | 0 \rangle = 1$. 2. Assuming $i$ is even, we have $\langle 0 | \gamma_i \gamma_{i+1} | 0 \rangle = i$ and $0$ for any other pair. 3. Assuming $i$ is odd, we have $\langle 0 | \gamma_i \gamma_{i-1} | 0 \rangle = -i$ and $0$ for any other pair. To conclude, we can solve and write all these expressions explicitly and numerically evaluate the parity.

Starting from another product state of the fermions involves only changing some of the correlators. For that, we only calculate the even case $\bra{1} \gamma_{2i} \gamma_{2i+1} \ket{1}$ (as for $i\neq j$ we will have unmatching fermion numbers on either the $i$'th site or the $j$'th site). The results can be calculated explicitly using the fermions notation $\bra{1} \gamma_{2i} \gamma_{2i+1} \ket{1} =\bra{1} (f_i + f_i^{\dagger}) (-i (f_i -f_i^{\dagger}))  \ket{1} = -i \bra{1} f_i^{\dagger} f_i  \ket{1} = -i$. Using anti-commutation relations we have that $\bra{1} \gamma_{2i} \gamma_{2i+1} \ket{1} = i$. The odd case is obtained by anti-commuting the even case. Therefore, we can calculate the correlations even for initial states as  $\ket{-++\dots++-}$ in the spin case.

\section{Parity switching  simulations}
\label{se:freeFermionResults}
In this appendix we provide additional results and analysis of the phase transition. We now apply free fermion methods to extract further results allowing to access the phase transition for large systems.

\begin{figure}
	\centering
	\begin{tabular}{c}
		(a) \\
		\includegraphics[width=\linewidth]{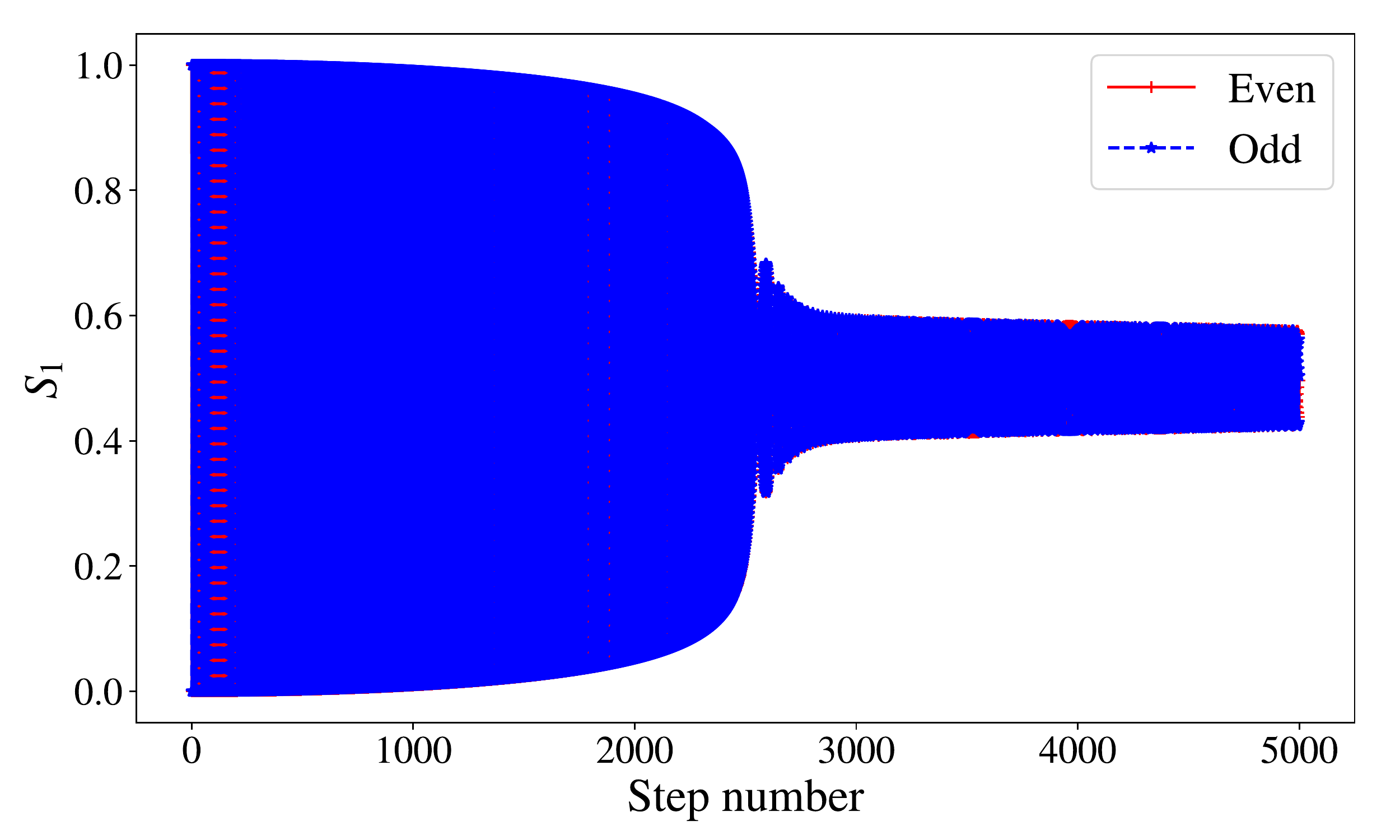} \\
		(b) \\
		\includegraphics[width=\linewidth]{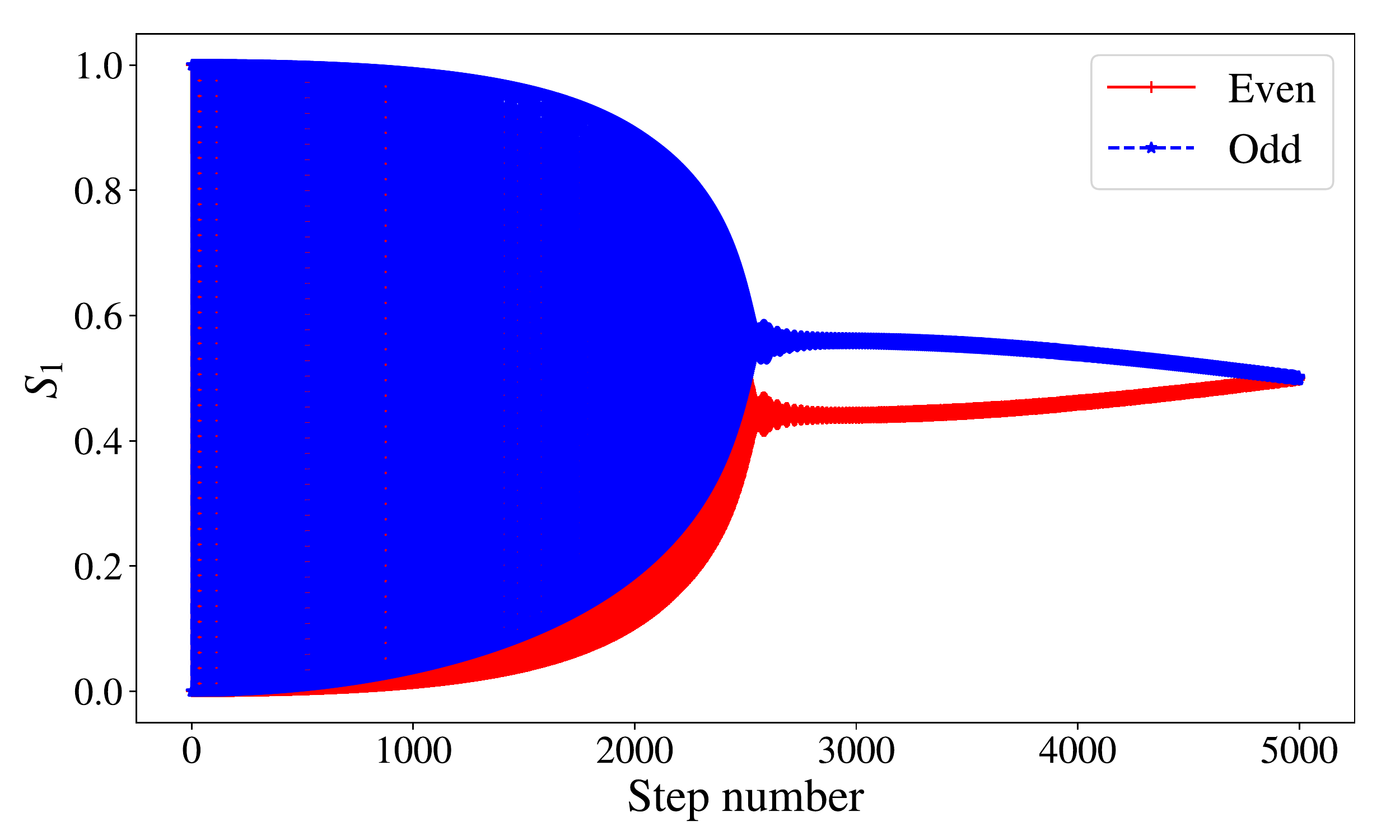} \\	
	\end{tabular}
	
	\caption{Calculations of symmetry resolved probabilities for  $L=100$ and $N_{\rm steps}=5000$ and either $L_A=50$ (a) or $L_A=1$ (b).}
	\label{fig:simulation_free}
\end{figure}

In Fig.~~\ref{fig:simulation_free}(a) we plot the subsystem parity probability $S_1$ for $L=100$ and $L_A=50$ and $N_{\rm steps}=5000$ steps. The parity indeed switches from even to odd as expected in the topological phase, and this switching decays very fast after the phase transition. Assigning a fine envelope function (see Fig.~2(c) in the main text) shows a first crossing at $S_1=0.5$ occuring at step number $2535 = N_{\rm steps} \times 0.507$. As we discuss below, we expect that as $L$ and $N_{\rm steps}$ grow the switching will decay almost completely in the trivial regime and the first crossing point will match the phase transition point at the step $0.5 \times N_{\rm steps}$.

The edge modes in the topological regime are expected to delocalize as we move towards the ferromagnetic phase, while at the start they are localized at the edge qubits. One can visualize the edge-delocalization effect by selecting a small subsystem size $L_A$, which here is taken to be $L_A=1$. The resulting $S_1$ is plotted in Fig.~~\ref{fig:simulation_free}(b). We observe that $S_1({\rm odd})$ and $S_1({\rm even})$ acquire different values close to the phase transition. This is in contrast to Fig.~~\ref{fig:simulation_free}(a), where we see that the probabilities of finding a large subsystem in opposite sectors is identical. This follows from the existence of a finite localization length $\xi$. Close to the transition, when $L_A \ll \xi $, the parity switching in the region $L_A$ is incomplete.

\begin{figure}
	\includegraphics[width=0.98\linewidth]{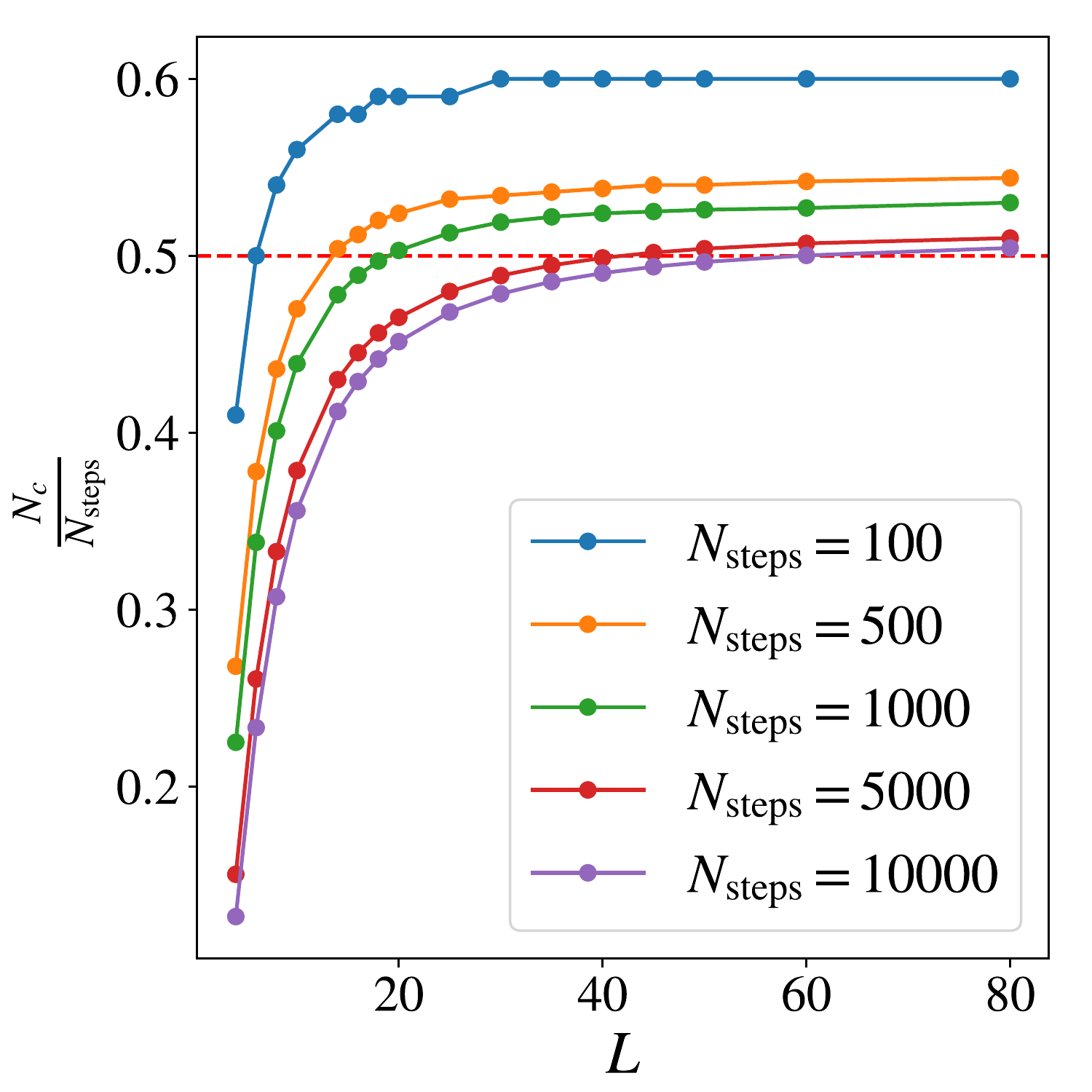} \\
	\caption{The crossing ratio $\frac{N_c}{N_{\rm steps}}$ for several  $N_{\rm steps}$ and varying system size $L$, where $N_c$ is the crossing step. This ratio converges to the phase transition point $N_c = 0.5 \times  N_{\rm steps}$ (dashed red line) in the limit $N_{\rm steps}\rightarrow \infty$, $L \rightarrow \infty$.}
	\label{fig:pt_10000}
\end{figure}

\subsubsection*{Phase transition point scaling}
Let us  focus on the case $L_A = L/2$ discussed in the main text. From Fig.~2(c) in the main text, one can operationally identify the crossing point of the envelope function with the phase transition. Using this definition, we are able to plot the step number of the crossing point, $N_c$, as a function of $L$ and $N_{\rm steps}$, see Fig.~\ref{fig:pt_10000}. We see that as $N_{\rm steps}$ increases, the curves converge in the large system size to $N_c = 0.5 N_{\rm steps}$, as expected. The mismatch for lower $N_{\rm steps}$ is understood from the accumulated phase of the not-exact $\pi$-mode as explained in Sec.~(\ref{app:2eigenstate}).

\section{Two Floquet eigenstates description of adiabatic evolution}
\label{app:2eigenstate}	
In this appendix we describe the two-Floquet eigenstate adiabatic evolution to characterize the phase transition from data such as Fig.~2(c) of the main text.

\subsection*{Envelope crossing in the vicinity of the phase transition}

As discussed in the main text, we see in Fig.~2(c) a clear signature of the transition near $\theta = \pi/4$ followed by a beating structure in $S_1({\rm{even}})$ in the ferromagnetic phase. We now explain this parity switching by using a two-Floquet eigenstate description.

An indication for the phase transition is seen from the quasi-energy spectrum of the Floquet operator $F$ along the adiabatic path shown in  Fig.~2(b) of the main text.
Starting from $F(\alpha,\beta) = F(1,\pi)$, the parameters $\alpha$, $\beta$ are changed continuously till the end point $\alpha=0$, $\beta=\pi-1$. For each such $F$ along the path, we construct the single-particle $2L \times 2L$ Majorana Hamiltonian $h$. Then, we plot for each such step along the adiabatic path the single-body quasi-energy spectrum of these $h$'s, which is defined from $-\pi$ till $\pi$. We see that at the phase transition point   we lose the Majorana $\pi$-mode (MPM) and are left only with the Majorana $0$-mode (MZM). We denoted by $\omega_{\pi-mode}$ the single particle quasi-energy corresponding to the $\pi-$mode in the topological phase (i.e. ${\rm{max}}(\omega_i)$ in the range $(-\pi,\pi)$).

Our initial state $| + \rangle$ is a linear superposition of two Floquet manybody eigenstates $|1\rangle$ and $|2\rangle$ with quasienergies $\omega_1$ and $\omega_2$ respectively, which differ by a $\pi-$single particle mode. The single-particle spectrum indicates the relative phase acquired by these two components upon Floquet evolution.

\begin{figure}[t]
	\includegraphics[width=\linewidth]{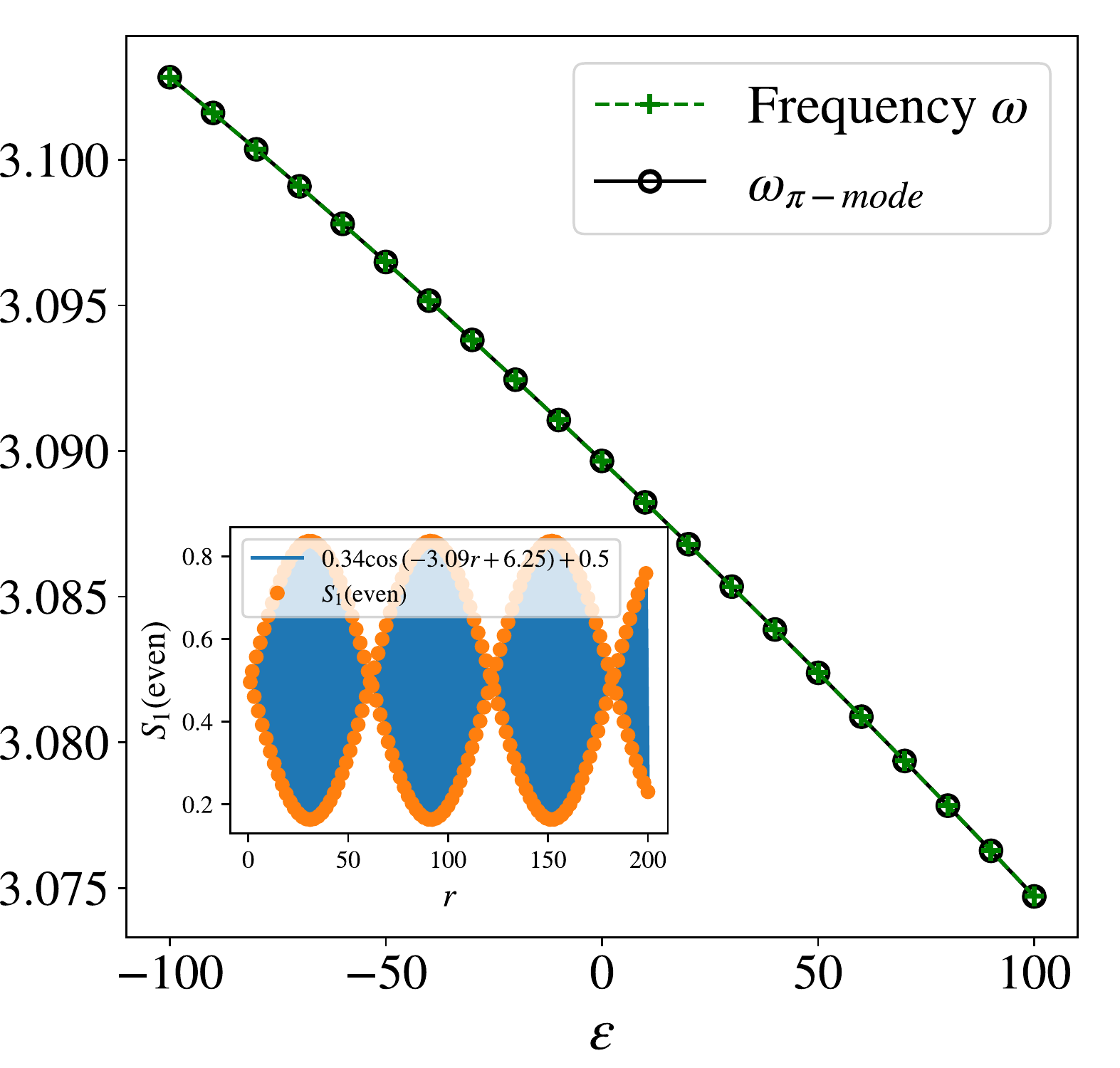}
	\caption{Fitting parameter $\omega$ defined in Eq.~(\ref{Aphiomega}) obtained via a calculation of $S_1({\rm{even}})$ for $L=2L_A=20$, $N_{\rm steps}=10000$ and $r=200$. We also plot the single particle quasi-energy $\omega_{\pi-mode}$ and observe that they coincide. Inset shows the fit for $\varepsilon=0$ stopping point, where the $x$ axis mark the $r$ repetitions.}
	\label{fig:fit_plot}
\end{figure}

\begin{figure}
	\includegraphics[width=\linewidth]{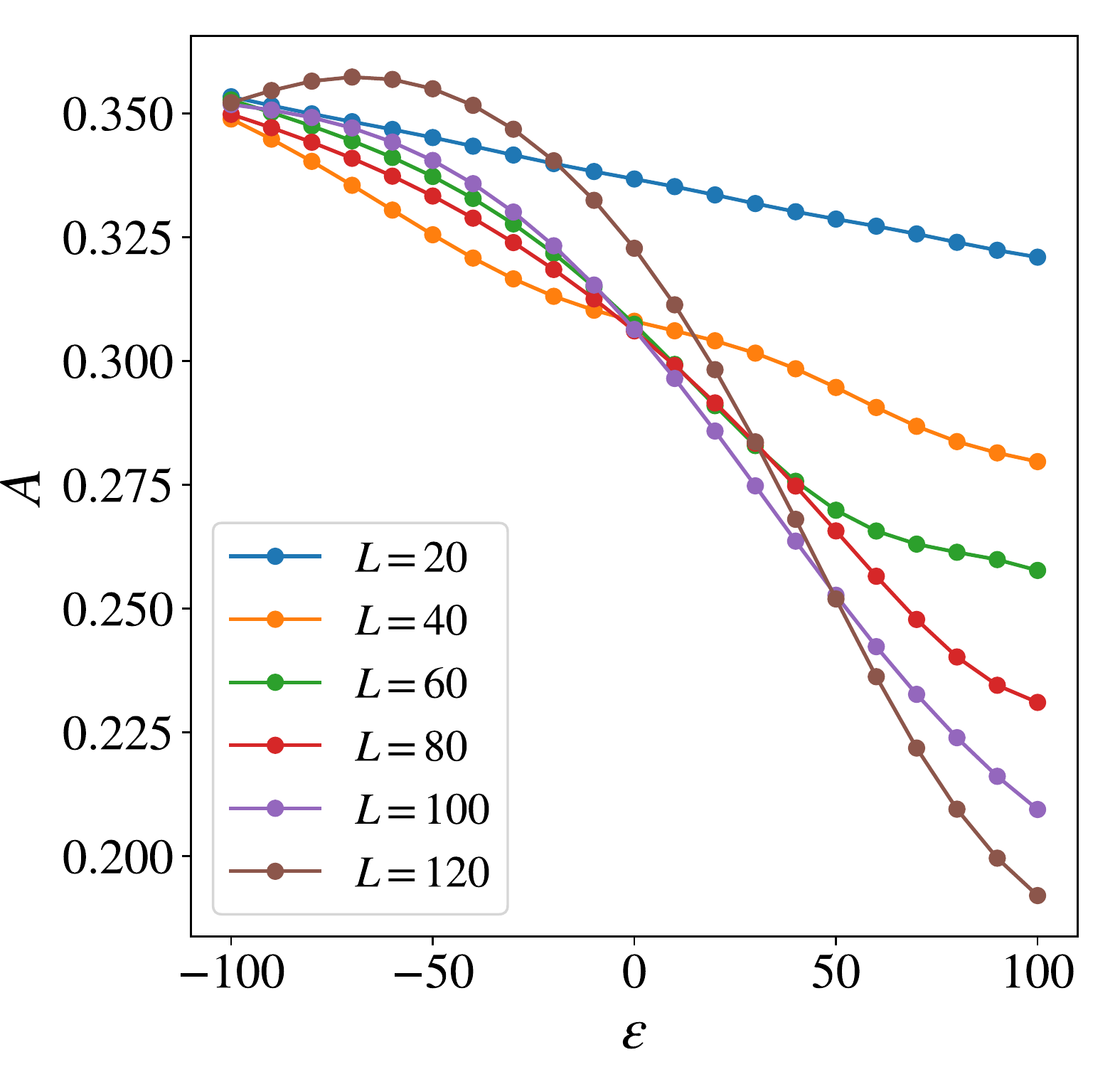}
	\caption{Fitting parameter $A$ defined in Eq.~(\ref{Aphiomega}) for several system sizes $L$, $L_A=L/2$, $N_{\rm steps}=10000$ and $r=50$. We see a  phase transition emerging for larger $L$'s.}
	\label{fig:fit_amp_plot}
\end{figure}

In order to describe a simple two Floquet eigenstates description, consider for simplicity the following protocol. We first proceed as in the adiabatic protocol described in the main text for $N_{\rm steps}/2+\varepsilon$ steps. Next, we evolve for $r$ repetitions according to a constant Floquet operator $F$ corresponding to the stopping point at the $N_{\rm steps}/2+\varepsilon$ step.

Now we examine the evolution of the subsystem parity along the $r$ last periods in the above protocol. Imagine starting from a state $| \Psi \rangle = a_1 |1\rangle + a_2 |2\rangle$ at some fixed value of $\theta$ corresponding to the stopping point, and then continue applying the same $F(\theta)$ for $0 \le n \le r$ times. Using the ansatz  $P_A |1 \rangle =e^{i \chi} | 2 \rangle$ for the topological phase, where $\chi$ is some phase, we see that 
\bea
\label{Aphiomega}
\langle  \Psi| P_A (n)|  \Psi \rangle &=& \langle  \Psi| {F^\dagger}^n P_A  F^n |  \Psi\rangle \nonumber \\ &=& a_1 a_2^* e^{in (\omega_2- \omega_1)}e^{i\eta}+ c.c. \nonumber \\
&=& 2A \cos( \omega n + \phi).
\eea
The last equation provides a simple fitting formula for the above protocol. Here, $\omega = \omega_2- \omega_1$  coincides with $\omega_{\pi-mode}$ at the stopping point.

In Fig.~\ref{fig:fit_plot} we plot the fit parameter $\omega$ near the transition, as a function of $\varepsilon =-100,-90,\dots,100$. To obtain this fit, we compute $S_1({\rm{even}})$ as function of $n$ and fit to the function $A \cos(\omega n + \phi)+0.5$ resulting from Eq.~(\ref{Aphiomega}). We then compare the fitting parameter $\omega$ to the single particle quasi-energy $\omega_{\pi-mode}$ and obtain an excellent agreement.

In Fig.~\ref{fig:fit_amp_plot} we plot the fitted amplitude $A$ near the transition for different system sizes. We can see that $A$ behaves like an order parameter of the FSPT phase, decaying from a finite value in the topological phase to near zero in the ferromagnetic phase. This decay sharpens as we increase the system size. Next we provide a theoretical description for this behavior.

\subsection*{Envelope beating in the ferromagnetic phase}

\begin{figure}
	\includegraphics[width=1.0\linewidth]{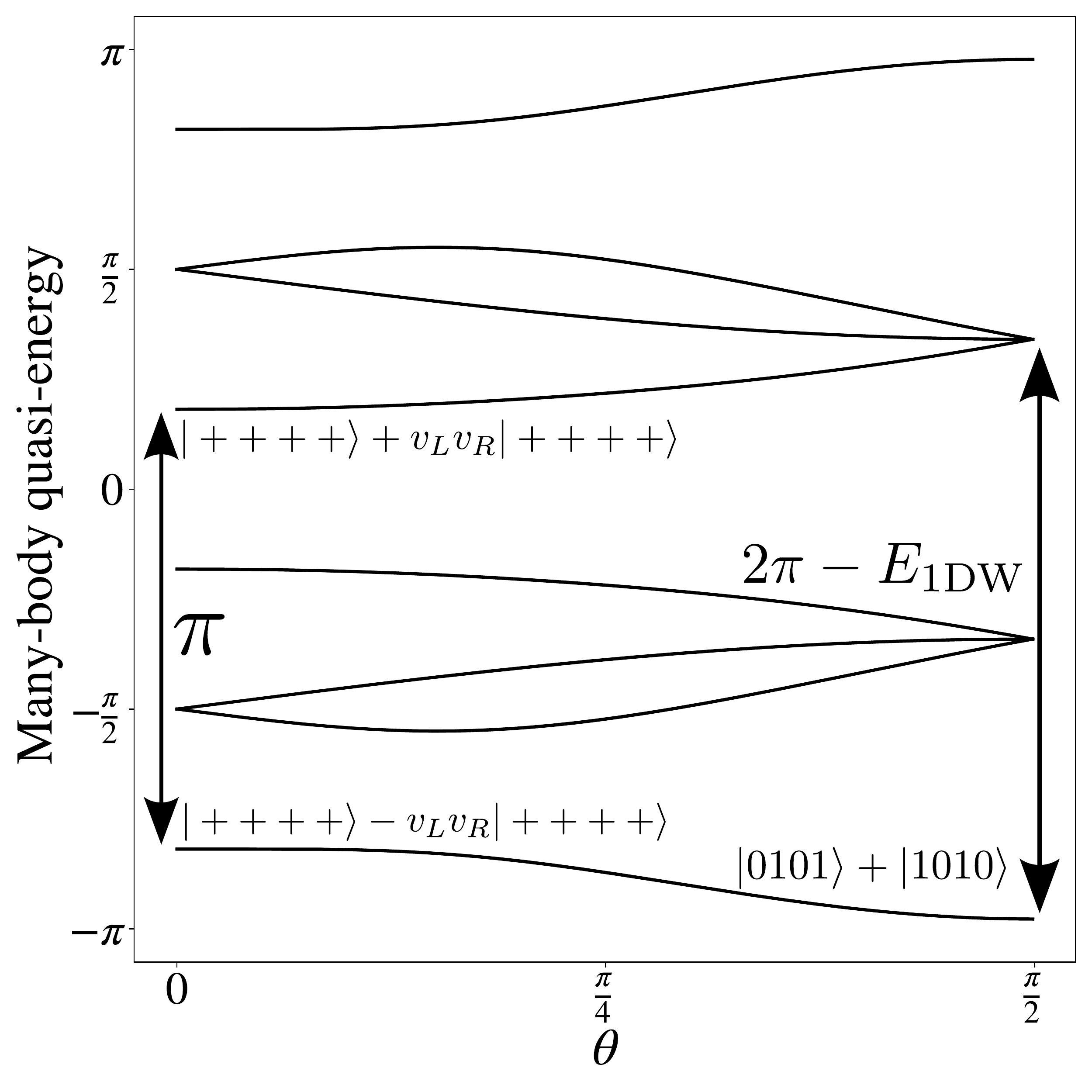}
	\caption{Many-body spectrum of the Floquet operator in Eq.~(7) of the main text for $L=4$. We identify the evolution of the initial $| + \rangle$ state as a linear combination of the two marked eigenstates. The quasienergy differnece changes from $\pi$ to $\beta$, the energy of a domain wall in the ferromagnetic phase.}
	\label{fig:mb_spec}
\end{figure}

The two Floquet eigenstate description persists all the way from the topological to the ferromagnetic phase. This holds as long as adiabaticity is maintained. To identify the relevant pair of Floquet eigenstates in the ferromagnetic phase we plot the many-body spectrum for 4 qubits as function of $\theta \in (0,\pi/2)$, see Fig.~\ref{fig:mb_spec}. Each level is two-fold degenerate due to the Majorana $0$-mode. The initial sate $|+\rangle $ at $\theta=0$ is a linear combination of two Floquet eigenstates
\bea
|1 \rangle &=& \ket{++++}+v_Lv_R\ket{++++}, \nonumber \\
|2 \rangle &=& \ket{++++}-v_Lv_R\ket{++++},
\eea
as marked on the figure. Here $v_L,v_R = e^{-i \frac{\alpha}{4} X_{L/R}} Z_{L/R} e^{i \frac{\alpha}{4} X_{L/R}}$ with $Z_L  = Z_1$ and $Z_R=Z_4$ and similarly for $X_{L,R}$. Up to a phase factor, $v_{L/R}$ flip the corresponding boundary spin from $| + \rangle$ to $| - \rangle$. We now follow this pair eigenstates to the ferromagnetic side $\theta = \pi/2$, where $F(\theta = \frac{\pi}{2}) = e^{-iH_{\rm FM}}$  with 
\be
H_{\rm FM} = -\frac{\beta}{2} \sum_{i=1}^{3} Z_i Z_{i+1}.
\ee

We can see in the figure that $|2 \rangle$ evolves into a state with quasienergy  $E_{0} = \frac{3\pi-3}{2} = \frac{3\beta}{2}$ (where $\beta =  \pi-r_0$ with $r_0=1$).
This state corresponds to the maximally excited state of the ferromagnet, where all bonds are anti-aligned (in the Floquet setting there is no notion of ground state, and adiabaticity only maps eigenstates to eigenstates). Similarly, the state $|1\rangle$ evolves to a state with energy $E_{1} =E_{0} - \beta = \frac{\beta}{2}$. This state is the second maximally excited state of the ferromagnet, having one less domain wall. The even parity state is a specific linear combination of $6$ 2-domain wall states, 
\be
\label{domainWallWaVEfUNCTION}
| E_{1} \rangle =(1+ \prod_{i=1}^4 X_i) (c_1 |0010 \rangle + c_2 |0110 \rangle + c_3 |0100 \rangle).
\ee

\begin{figure}
	\includegraphics[width=1.1\linewidth]{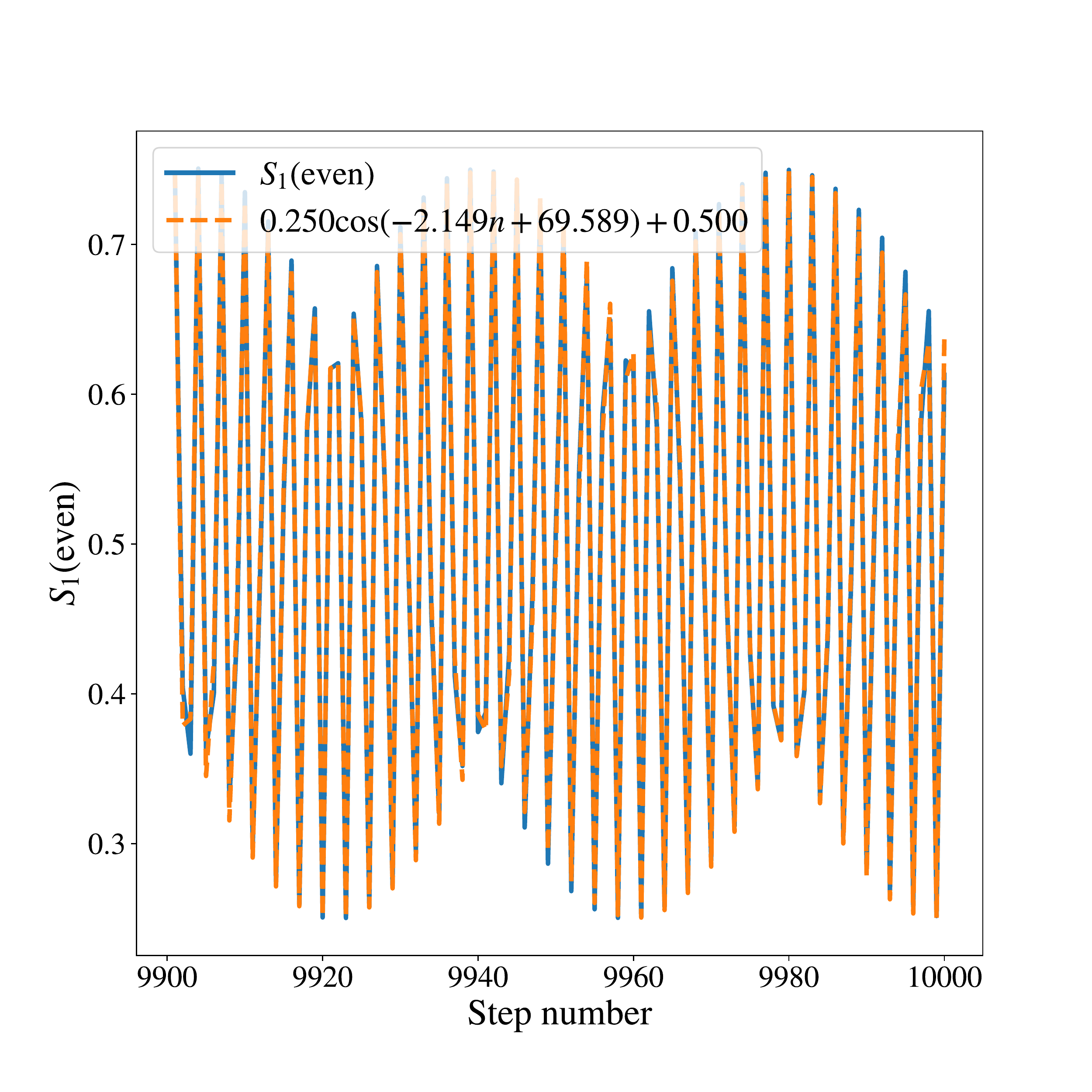}
	\caption{Symmetry resolved even probability for $L=2L_A = 8$ and $N_{\rm steps}=10000$, with a fit for the last $100$ points in the ferromagnetic phase with $A \cos{(\omega n + \phi)} + 0.5$.}
	\label{fig:fit_trivial}
\end{figure}

Here $c_i$ is the wave function amplitude to find that there is no domain wall in the link connecting the sites $i$ and $i+1$. Thus, by exact diagonalization of a small system we see that the $\pi$-quasienergy difference between the two initial Floquet eigenstates, changes in the ferromagnetic phase into the energy of a domain wall excitation $E_{\rm 1DW} = \beta$. Since this energy is not locked to a fraction of $2 \pi$, it leads to a beating structure. A fit of the form of Eq.~(\ref{Aphiomega}) deep in the ferromagnetic phase is shown in Fig.~\ref{fig:fit_trivial}. We see that $\omega \approx -2.149 \approx -\beta = 1-\pi$ as expected from the domain wall energy excitation.

\begin{figure}[t]
	\centering
	\begin{tabular}{c}
		(a) \\
		\includegraphics[width=\linewidth]{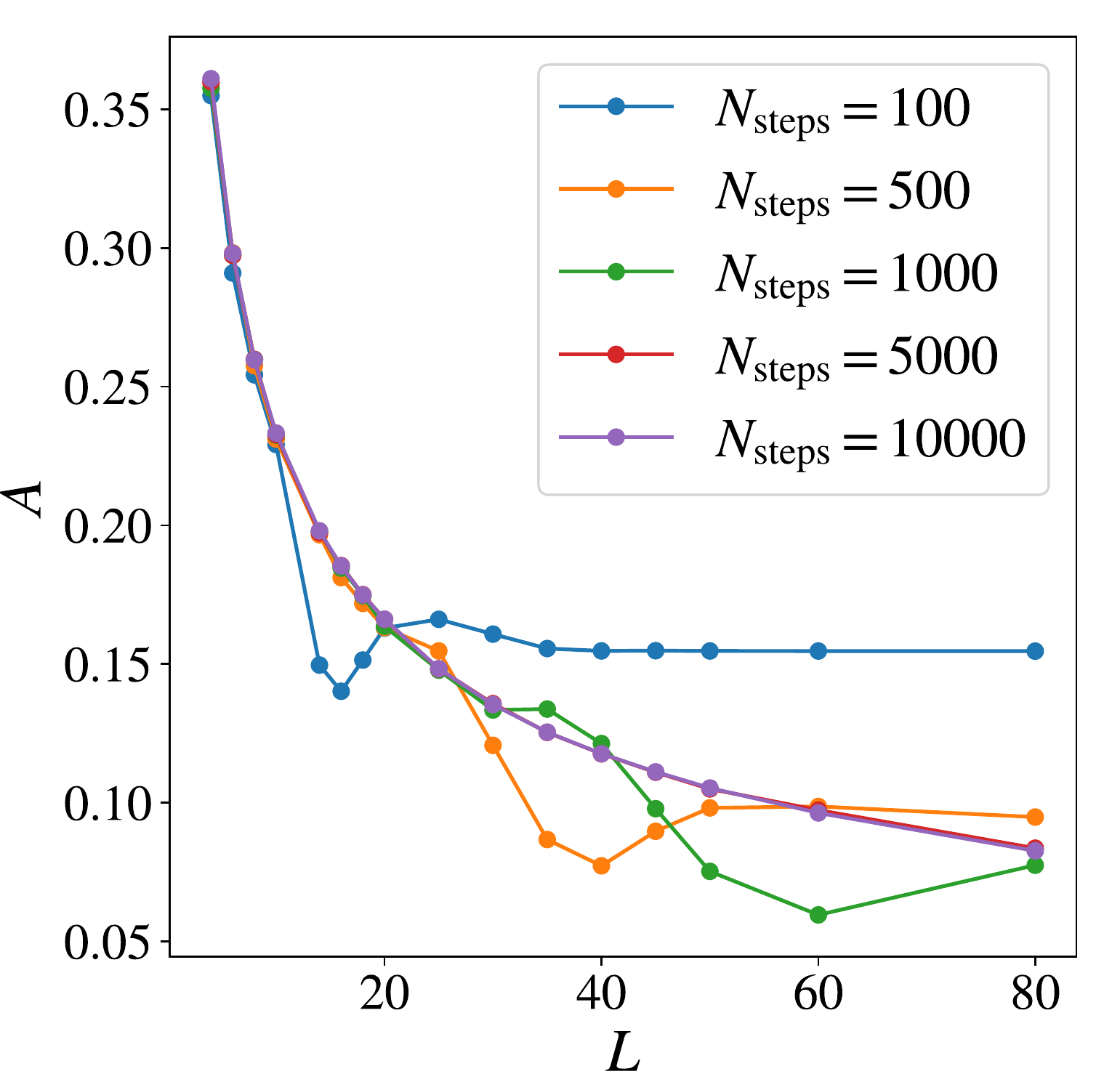}\\
		(b) \\
		\includegraphics[width=\linewidth]{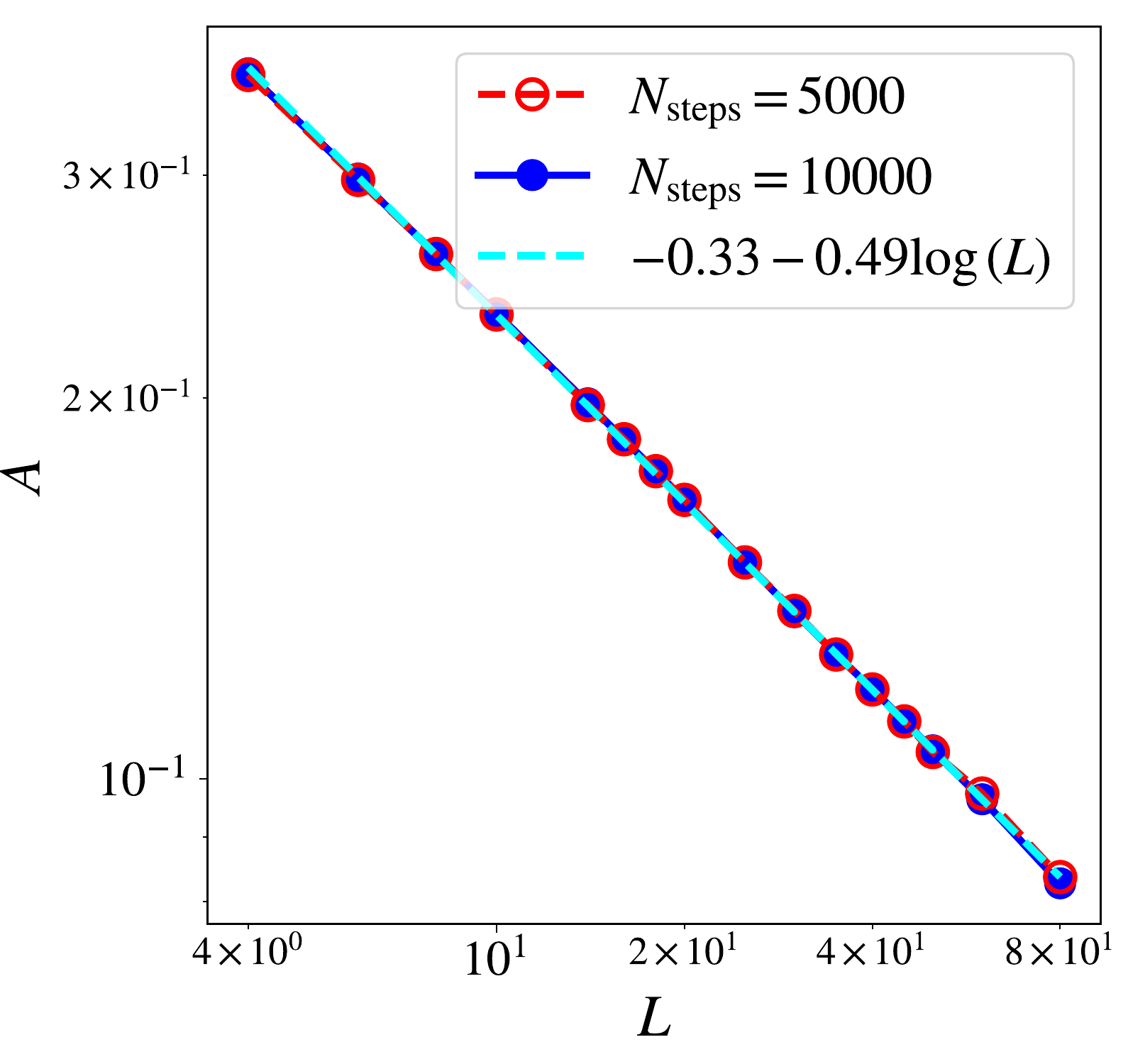} \\	
	\end{tabular}
	\caption{(a) Fit amplitude $A$ for the ferromagnetic regime (each amplitude is generated from a fit with $\frac{N_{\rm steps}}{10}$) as function of $L$ for various $N_{\rm steps}$. (b) In the adiabatic limit $A$ follows a $ \frac{1}{\sqrt{L}}$ behavior.}
	\label{fig:fit_trivial_amps}
\end{figure}

\begin{figure}[t]
	\includegraphics[width=\linewidth]{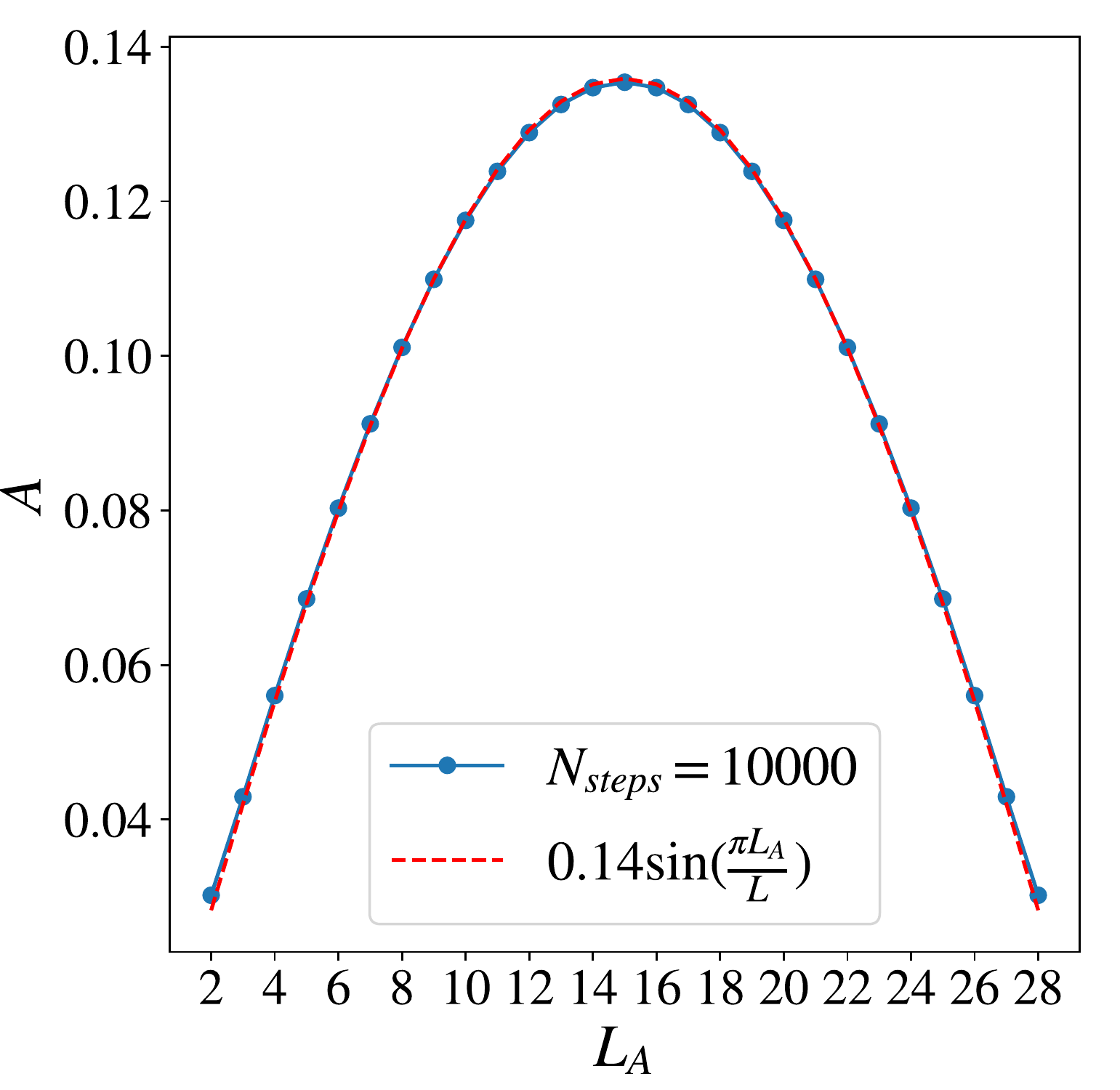}
	\caption{Amplitude fit $A$ as a function of the subsystem size $L_A$ for $L=30$ accompanied with a dashed red line fit of the  ground state wave-function for particle in an infinite potential well.}
	\label{fig:fit_trivial_amp_subsys}
\end{figure}

Valuable information can be extracted by examining the dependence of the fit amplitude $A$ on $L$ and $N_{\rm steps}$. This is plotted in Fig.~\ref{fig:fit_trivial_amps}(a). We see that $A$ decreases as $L\rightarrow \infty$. Moreover, we see the importance of the adiabatic limit in seeing this transition. In Fig.~\ref{fig:fit_trivial_amps}(b) we show a logarithmic plot of the results in the adiabatic limit, which shows the relation $A \propto \frac{1}{\sqrt{L}}$. We next explain the origin of this relation.

We revise the argument in Eq.~(\ref{Aphiomega}) in terms of the two eigenstates in the ferromagnetic phase. The matrix element of $P_A$ between the two eigenstates in the ferromagnetic phase, consisting of the ferromagnetic (maximally excited) state and a state differing from it by a single-domain wall excitation, is proportional to the amplitude $c_{L_A}$ of the domain wall  to be located at the interface between the two subsystems, see Eq.~(\ref{domainWallWaVEfUNCTION}). To determine this wave function, 
in Fig~\ref{fig:fit_trivial_amp_subsys} we plot the amplitude $A$ fit for the trivial regime for each bipartition subsystem of size $L_A$ (excluding the edge qubits) in  a $L=30$ system. We observe a match with the groundstate of a particle in an infinite well of size $L$, which is $c_j \propto \sin(\pi \frac{j}{L})$. From the normalization of this wave function, the amplitude for half of the system $L_A=\frac{L}{2}$ scales as $\frac{1}{\sqrt{L}}$.

\begin{figure*}[t]	
	\centering
	\begin{tabular}{c c}
		(a) Qubit-resolved parity & (b) SRE switching \\
		\includegraphics[width=0.5\linewidth]{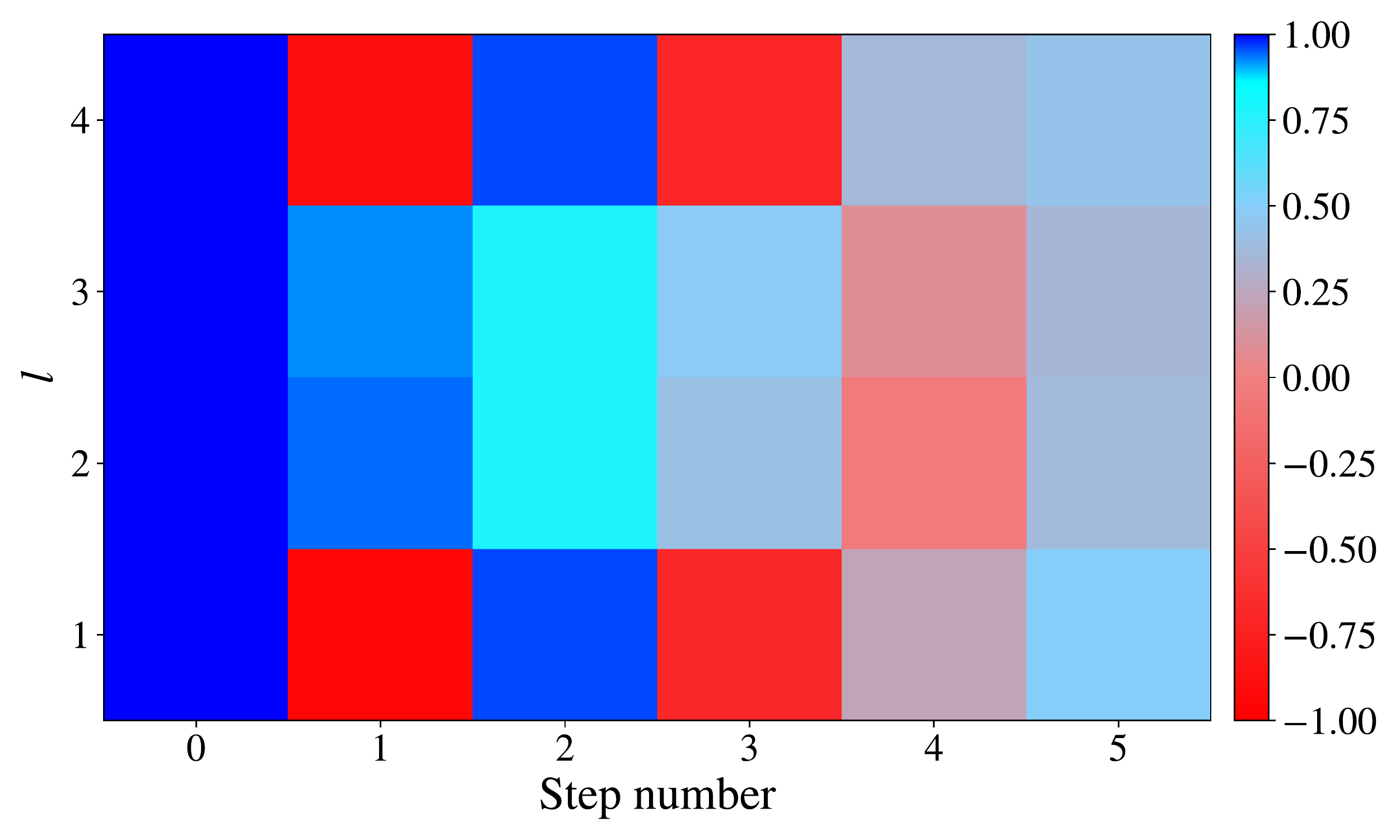} &
		\includegraphics[width=0.5\linewidth]{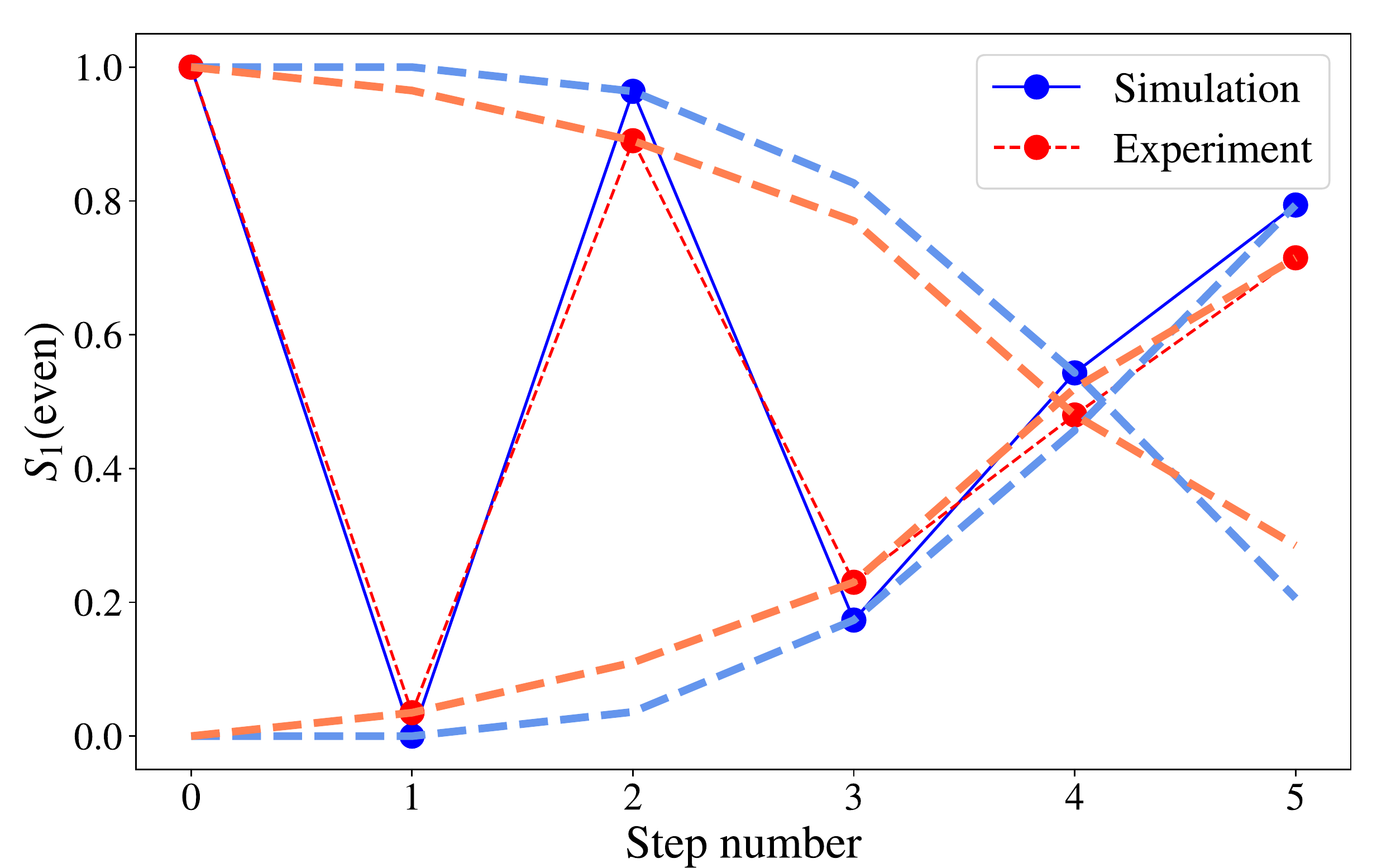}	\\
		\includegraphics[width=0.5\linewidth]{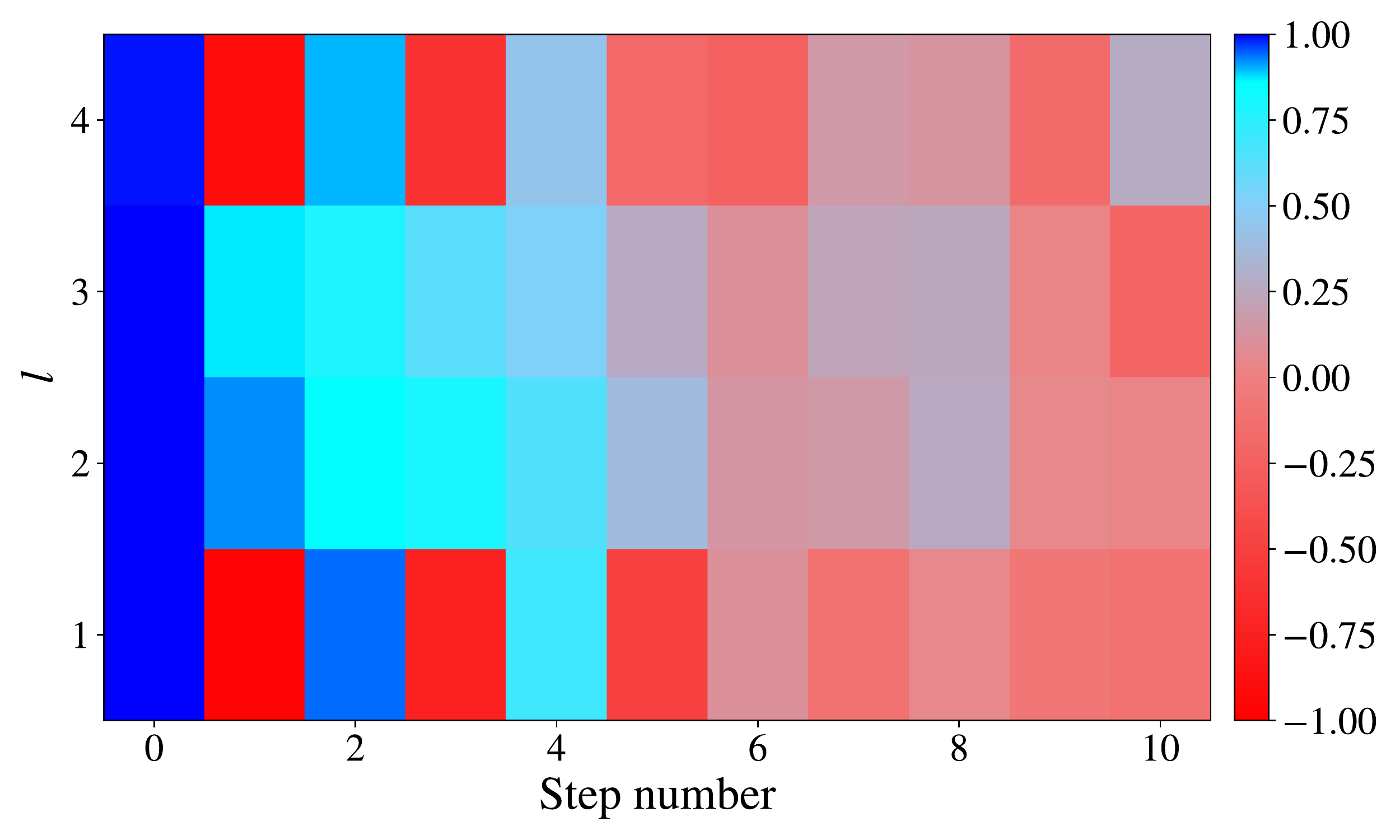} &
		\includegraphics[width=0.5\linewidth]{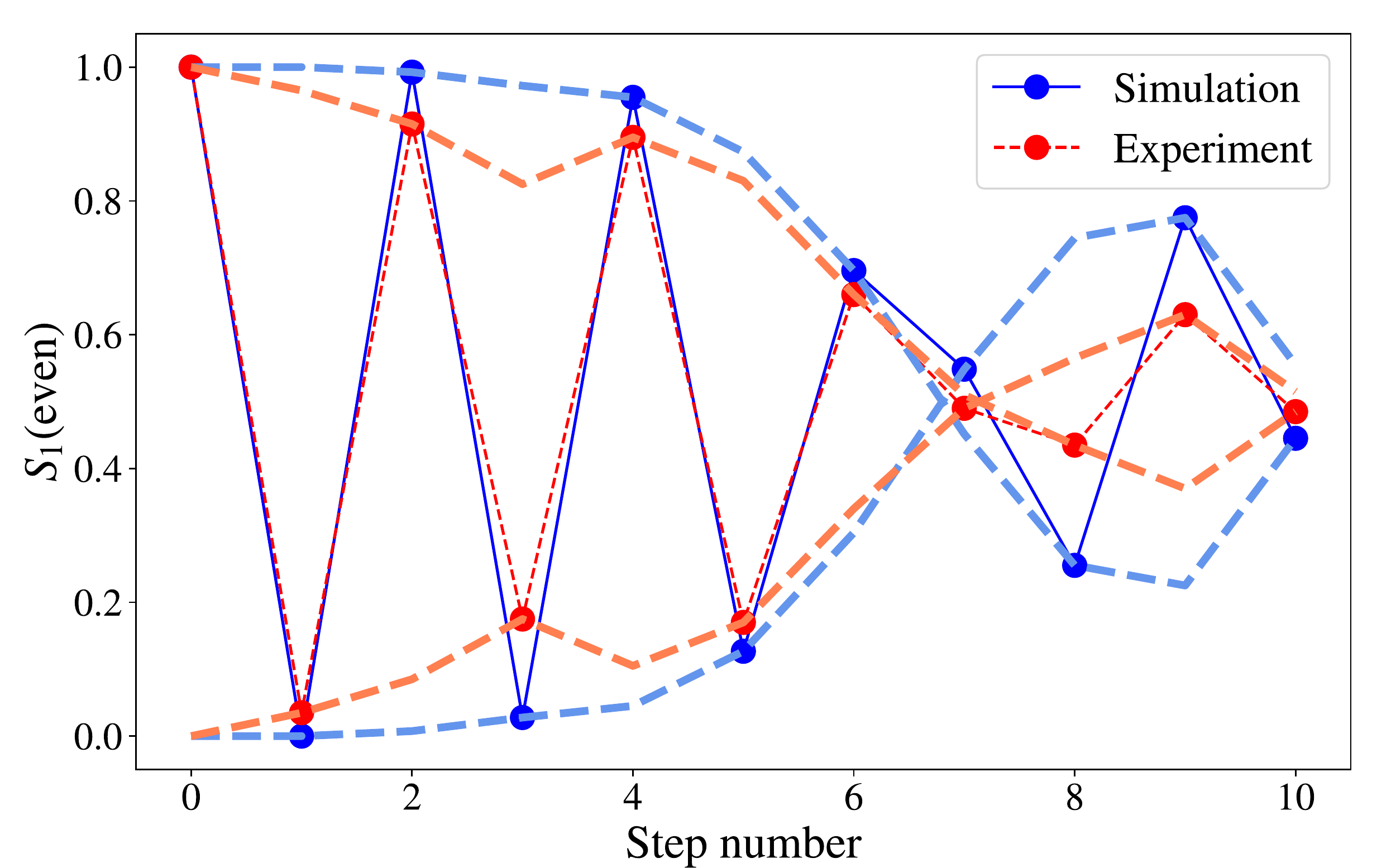} \\
		\includegraphics[width=0.5\linewidth]{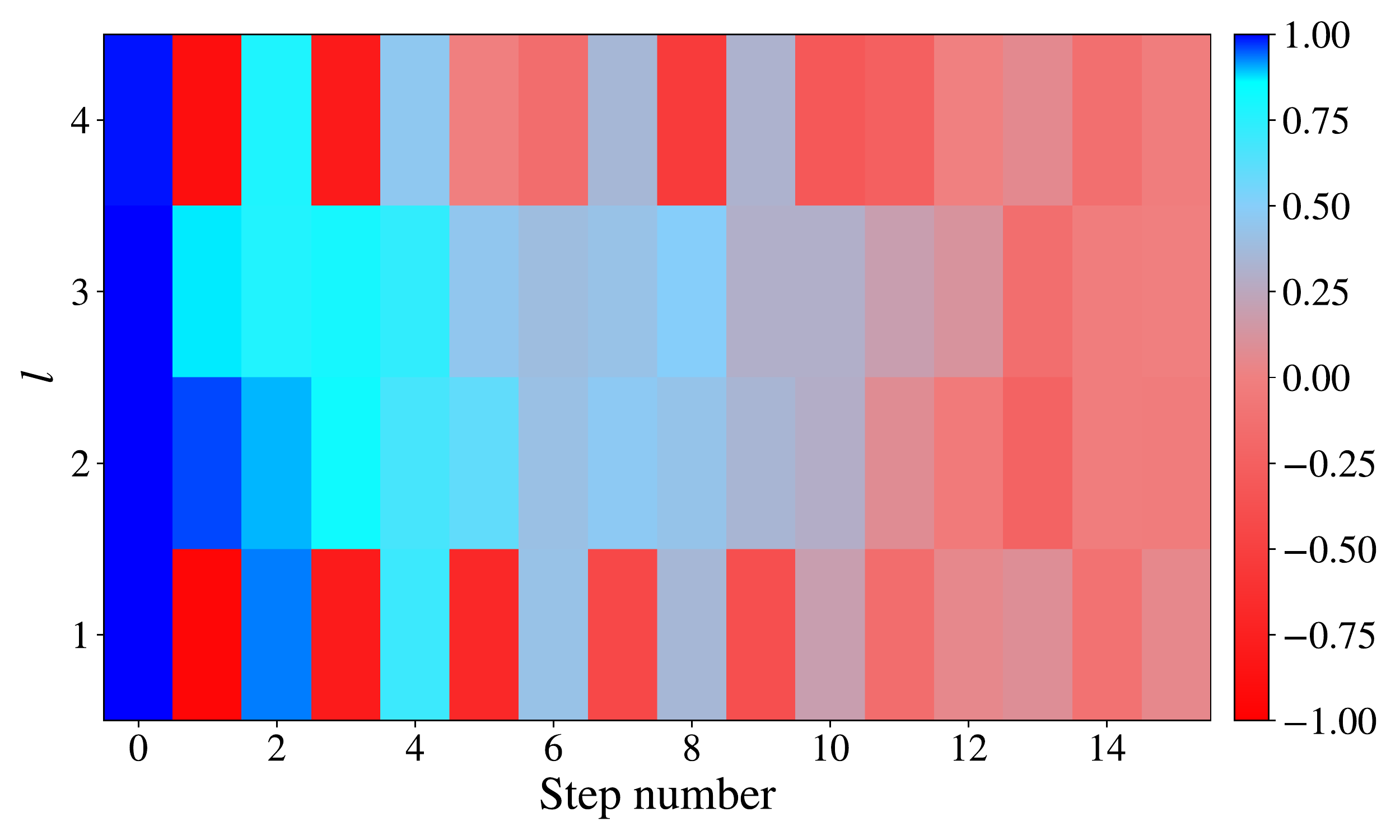} &
		\includegraphics[width=0.5\linewidth]{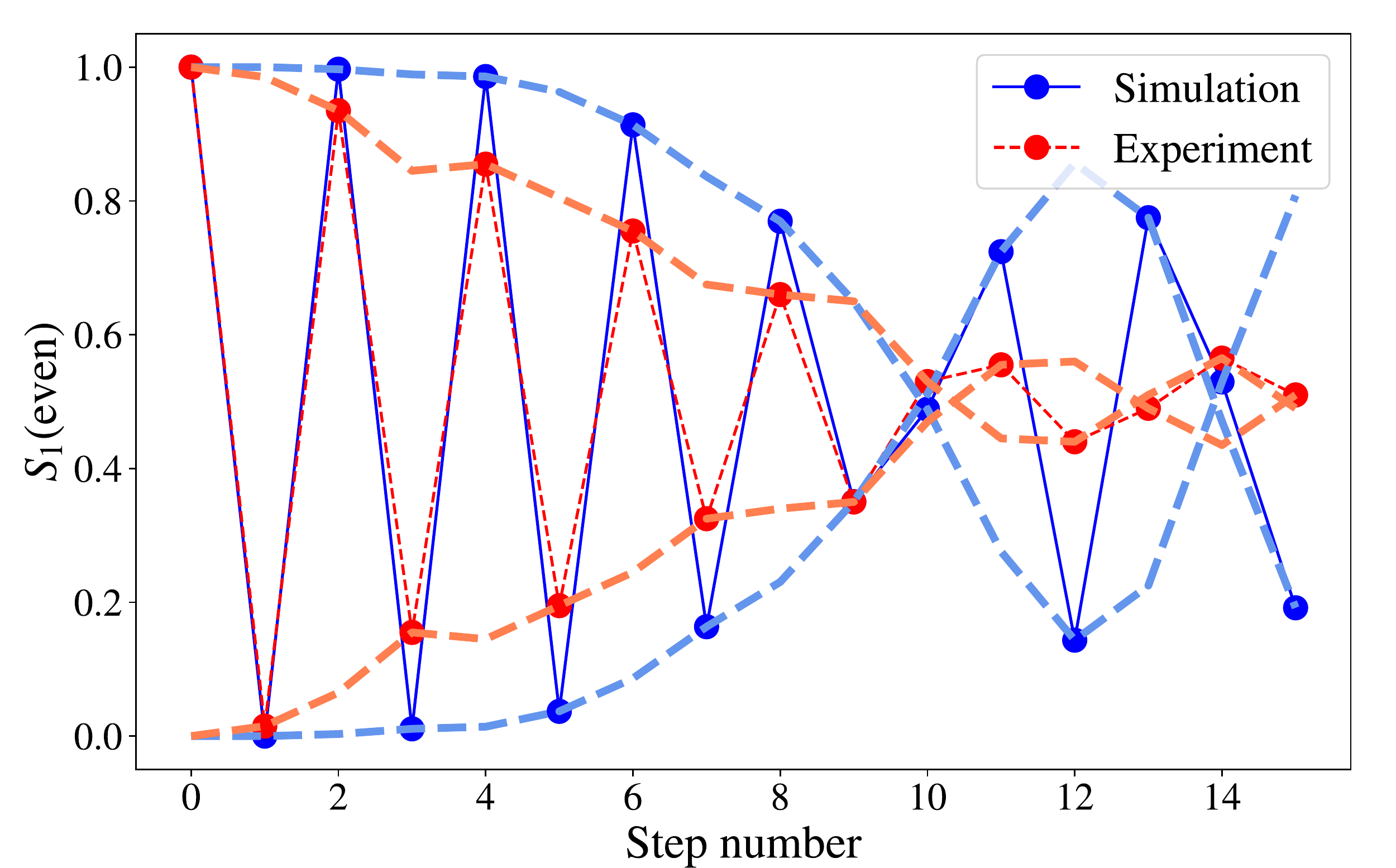} 
	\end{tabular}
	\caption{Adiabatically crossing a topological phase transition on the IonQ NISQ computer. We use $L=4$ qubits  for several $N_{\rm steps}=5,10,15$. (a) Qubit-resolved parity across the adiabatic path, where for the $l$'th qubit we have color plotted the $\langle X_l \rangle$. (b) Plot of the experimental $S_1({\rm{even}})$ along with the theoretical prediction. An envelope was plotted to mark the crossing, which occurs when the two lines of the envelope intersect. }
	\label{fig:exp_results}
\end{figure*}

\section{Experimental results on NISQ computer}	

We provide all the experimental runs for different system sizes $L$ and $N_{\rm steps}$. We have performed all the experiments on the IonQ NISQ computer, which is an ion-trapping based computer with $11$ qubits. In Fig.~\ref{fig:exp_results} we see that the experimental and theoretical predictions of the switching are in a very good agreement. The phase transition can also be seen clearly from the color plots, where the blue color turns to red. A python code example for the results is presented below.

\onecolumngrid

\pythonstyle
\begin{lstlisting}[label = {alg:fspt}, caption = {Python Code}]
# general imports
import matplotlib.pyplot as plt
# magic word for producing visualizations in notebook
%matplotlib inline
import string
import time
import numpy as np
import random as rd

import json
import os


# AWS imports: Import Braket SDK modules
from braket.circuits import Circuit, Gate, Observable
from braket.devices import LocalSimulator
from braket.aws import AwsDevice, AwsQuantumTask



def add_one_step(qc,Nqubits,alpha,beta) :
    for i in range(Nqubits) :
        qc.rx(i,-alpha)
    for i in range(Nqubits-1) :
        qc.cnot(i,i+1)
        qc.rz(i+1,-beta)
        qc.cnot(i,i+1)

#used to prepare the g.s. of the sweet point and to measure X
def add_rotate_all(qc,Nqbits) :
    for i in range(Nqubits) :
        qc.h(i)

def average(items,Nqubits) :
    average=np.zeros(Nqubits)
    for state, count in items :
        value=np.zeros(Nqubits)
        for i in range(Nqubits):
            value[i]=(1-2*int(state[i]))
        average+=count*(value)
    return average

def parity(items,cut):
    parity=0;
    for state, count in items :
        value=1
        for i in range(cut):
            value=value*(1-2*int(state[i]))
        parity+=value*count
    return parity

#PHYSICAL PARAMETERS

Nshots=1000;
Nqubits=4;
Nsteps=10;
radius=1.0;


device = LocalSimulator(); 
#TO run on actual device use
#device = AwsDevice("arn:aws:braket:::device/qpu/ionq/ionQdevice");

#BUILDING AND RUNNING THE CIRCUIT

qc=Circuit()

#prepare the g.s. of the sweet point
add_rotate_all(qc,Nqubits)

angles=np.pi/2*np.array(range(Nsteps))/(Nsteps-1)
alphas=radius*np.cos(angles)
betas=np.pi-radius*np.sin(angles)

all_id=[] #used for AWS machines
all_prob=[] #used for local simulator

for n in range(Nsteps+1):
    
    if n>0 :
        add_one_step(qc,Nqubits,alphas[n-1],betas[n-1])

    #rotate in X direction
    temp_qc=qc.copy()
    add_rotate_all(temp_qc,Nqubits)
    
    #run and measure
    #for actual device add "s3_destination_folder"
    all_prob.append(device.run(temp_qc,shots=Nshots).result().measurement_probabilities)
    
#COMPUTING EXPECTATION VALUES

all_avg=[]
all_parity=[]


for n in range(Nsteps+1): 
    items=all_prob[n].items()
    all_parity.append(parity(items,int(Nqubits/2))); 
    all_avg.append(average(items,Nqubits))
    

print("N_qubits = "+str(Nqubits))
print("time steps = "+str(Nsteps))
print("radius = "+str(radius))

print("average parity of L/2 qubits")
print(all_parity)
print("polarization")
print(np.array(all_avg))

# RESULTS:
# N_qubits = 4
# time steps = 10
# radius = 1.0
# average parity of L/2 qubits
# [1.0, -1.0, 0.98, -0.9640000000000001, 0.9059999999999999, -0.73, 0.35999999999999993, 0.057999999999999975, -0.476, 0.5760000000000002, -0.10199999999999998]
# polarization
# [[ 1.     1.     1.     1.   ]
# [-1.     1.     1.    -1.   ]
# [ 0.99   0.97   0.97   0.99 ]
# [-0.914  0.882  0.922 -0.95 ]
# [ 0.688  0.686  0.664  0.706]
# [-0.618  0.592  0.582 -0.592]
# [ 0.114  0.318  0.306  0.118]
# [-0.014  0.528  0.468  0.03 ]
# [-0.372  0.552  0.564 -0.34 ]
# [ 0.362  0.302  0.344  0.432]
# [-0.062  0.292  0.27  -0.048]]
\end{lstlisting}

\twocolumngrid

\section{Teleportation from FSPT algebra}
In the main text we demonstrated the teleportation protocol for a specific point in the FSPT phase. Here we formulate the teleportation procedure for $G=\intz_2$ only based on the general algebra in Eqs.~(3) and (4) of the main text. Hence the ability to teleport using the controlled-$F$ gate, is a property of the entire phase.

Start from a Floquet eigenstate in the bulk, such that $e^{-if}$ generates just a phase. We encode a ``qubit" $\ket{\psi} = \sum_{g} \alpha_g | g \rangle$  
at the left edge %(arbitrary edge choice) 
in the $G$-symmetry eigenbasis $|g \rangle = |\pm \rangle$ and we encode the identity group element state $|+\rangle$ in the right-edge qubit. We denote the initial state in terms of the left and right edges $| \psi \rangle_L \otimes | + \rangle_R $.  The action of $F$ on the bulk qubits generates a phase $e^{-i f}$. Applying $I+F  e^{i f} $ we reach the state $ \left[ I+v_Lv_R \right]\ket{\psi}_L \otimes \ket{+}_R$. Next, we measure the left qubit and assume it is found in the $| + \rangle$  state. Using Eq.~(6) of the main text, we have  $\Pi_+ v_L = v_L \Pi_-$. Hence, 
\bea \Pi_+ \left[ I+v_Lv_R \right]\ket{\psi}_L \otimes \ket{+}_R =\nonumber  \\\left[ \alpha_{+}  + \alpha_{-} \bra{+}v_L\ket{-}  v_R  \right] \ket{+}_L \otimes \ket{+}_R 
.
\eea 
The left qubit has been measured in $|+ \rangle$ and the state of the right qubit becomes
\bea    \alpha_{+} \ket{+} &+& \alpha_{-} \bra{+}v_L\ket{-} v_R \ket{+}  =\nonumber  \\   \alpha_{+} \ket{+} &+& \alpha_{-} \bra{+}v_L\ket{-} \bra{-} v_R \ket{+} \ket{-} 
.
\eea
We can see teleportation up to a nonuniversal phase factor $\bra{+}v_L\ket{-}\bra{-}v_R\ket{+}$. This phase cancels in the case that $v_R$ is locally the same as $v_L$, generating the opposite phase. We emphasize that this teleportation from the left follows only from the underlying FSPT algebra. For a trivial FSPT, $v_L$ commutes with $\Pi_+$ and there is no teleportation. Similarly, teleportation of an $N-$qudit can be shown in a similar way for the more general case with $G=\intz_N$; on of the modifications in the case, is to need to apply $\sum_{k=0}^{N-1} (F e^{if})^k$ rather than $1+e^{if}F$.

\section{FSPT action on the signature of general finite Abelian group $G$}

In this section we  prove general results on the degeneracy of  symmetry sectors. The main result is Eq.~(\ref{eq:mainresult}) which describes symmetry sectors cycling upon Floquet  evolution in the presence of degeneracies due to nontrivial static SPT order. We begin by providing a mathematical description of these degeneracies in the static case.

\subsection*{SPT degeneracies}

Our main quantity of interest is the list of length $|G|$ of symmetry resolved probabilities $\{ S_1(\calq) \}$ whose degeneracies classify the SPT phase. We define the signature $(s_1,s_2 , \dots)$ of this list such that the $i$'th element of the signature $s_i$ is the degeneracy of the $i-th$ element (choosing any order) of $\{ S_1(\calq) \}$ after removing from it degenerate elements. For example, the signature of the list $\{ 3,3,5,5,6 \}$
is $(1,2,2)$ (the order here is arbitrary).

Our approach to capture the degeneracies is by looking for non zeros of $Z_g = \Tr[U_A(g)\rho]$ based on Ref.~\cite{azses2020symmetry}. Here we considered pure states $\rho$ describing SPT fixed points build in terms of cocycles according to Ref.~\cite{chen2013symmetry}, and $Z_g$ receives an interpretation as the partition function in the presence of a defect separating the subsystems and carrying a group element $g$.  Let us assume (this is provable for the case of $\intz_N \times \intz_N$) that the $g$'s such that $Z_g \neq 0$ form a group, which we call $H \leq G$. Therefore, we can simplify the following formula from Ref.~\cite{azses2020symmetry} (see Eq.~5)
\be
S_1(\calq) = \frac{1}{|G|} \sum_{g \in G} \chi_{\calq}(g) Z_g.
\ee
As $g$'s such that $Z_g=0$ contribute nothing to the sum, we can restrict the sum to those $g$'s with non-zero $Z_g \neq 0$,
\be
S_1(\calq) = \frac{1}{|G|} \sum_{g \in H} \chi_{\calq}(g) Z_g.
\ee
We notice that if two symmetry sectors $\calq_1,\calq_2$ have their character fixed on $H$, defined as $\chi_{\calq_1}(g)=\chi_{\calq_2}(g)$ for every $g \in H$, then $S_1(\calq_1) = S_1(\calq_2)$.

In parallel to the definition of signature, we define families $F_{\calq}$. Families are subsets of $G$ such that every group element is in one and only one family (a.k.a. partition of $G$). Two group elements are in the same family based on their respective character fixed values over $H$. Group elements $\calq_1,\calq_2 \in G$ are in the same family if and only if $\chi_{Q_1}(g)=\chi_{Q_2}(g)$ for every $g \in H$. The family $F_\calq$ contains the group element $\calq$ (there are multiple choices as every element in the family $F_\calq$ can be chosen). For an Abelian group $G$, the number of characters that are fixed over a subgroup $H \leq G$ is $[G:H]$, the index of $H$ in $G$, which for finite groups is $\frac{|G|}{|H|}$ \cite{berkovich1998characters} (see pp. 15). Hence, in each family there are exactly $\frac{|G|}{|H|}$ characters fixed over $H$, creating a degenerate sector in the symmetry-resolved probabilities $S_1(\calq)$. This degeneracy is doubled for $F_\calq \neq F_{\calq^{-1}}$, as $S_1(\calq^{-1}) = S_1(\calq)$, which we prove below.

The different families of $G$ fix the signature. Assuming no degeneracy doubling between different families, the signature is 
\be
(\underbrace{|G|/|H|}_{ |H| ~{\rm{times}}}).
\ee 
This applies to a limited set of cases, for example for $H=\intz_2 \times \intz_2$ and any finite Abelian group $G$ with $H \leq G$, and we obtain a signature of $(|G|/4,|G|/4,|G|/4,|G|/4)$. Alternatively for $G=\intz_2 \times \intz_2$ and $H=\intz_1$ we have signature (4), corresponding to the maximal fourfold degeneracy.

Let us prove first that $Z_g$ is real for finite Abelian group $G$ based on cocycles properties and Ref.~\cite{azses2020symmetry}. First, there is a choice of cocycles (see Refs.~\cite{azses2020symmetry,chen2013symmetry} for definitions) such that $\omega(ab,c) = \omega(a,c)\omega(b,c), \ \omega(a,bc) = \omega(a,b)\omega(a,c)$ for $a,b,c \in G$, where $G = \intz_{e_1} \times \dots \times \intz_{e_{l}}$ is a general finite Abelian group with $e_i$ divides $e_{i+1}$ (such a decomposition always exists as a result of the classification of finite Abelian groups). The proof is based on the usual choice $\omega(a,b) = \exp{(2\pi i \sum_{i<j} \frac{p_{ij} a_ib_j}{d_{ij}})}$, where $a_i \in \intz_{e_i}$ are representing the element $a \in G$, $d_{ij} = {\rm gcd}(e_i,e_j)$ and $p_{i<j}$ enumerates the different SPT phases ($p_{i>j}=0$ for convenient). The claim is now proved as $\omega(ab,c) = \exp{(2\pi i \sum_{i<j} \frac{p_{ij} (a+b)_ic_j}{d_{ij}})} =  \exp{(2\pi i \sum_{i<j} \frac{p_{ij} a_ic_j}{d_{ij}})} \exp{(2\pi i \sum_{i<j} \frac{p_{ij} b_ic_j}{d_{ij}})} = \omega(a,c)\omega(b,c)$, and similarly for the other identity. We write Eq.~(21) in Ref.~\cite{azses2020symmetry}
\bea
&&Z_g= \nonumber \\
&&\frac{1}{|G|^3} \left[ \sum_{s_1} \frac{\beta(s_1g)}{\omega(g,s_1)\beta(s_1)} \right] \left[ \sum_{s_2} \frac{\omega(s_2,g)\beta(s_2)}{\beta(s_2g)} \right] f(g),\nonumber \\
\eea
where $f(g) = \left[ \sum_{s_3} \frac{\omega(g,s_3)}{\omega(s_3,g)}\right] $ and $\beta(g)$ is a coboundary (arbitrary function from $G$ to $U(1)$). Combining $f(g)$ and the second sum and changing variables $s_3 \rightarrow s_3s_2$ we transform the inner sum to $\sum_{s_3} \frac{\omega(g,s_2s_3)}{\omega(s_2s_3,g)} = \frac{\omega(g,s_2)}{\omega(s_2,g)} \sum_{s_3}\frac{\omega(g,s_3)}{\omega(s_3,g)} = \frac{\omega(g,s_2)}{\omega(s_2,g)} f(g)$. Therefore, we have 
\bea
Z_g=\frac{1}{|G|^3} \eta(g) \overline{\eta(g)} f(g),\nonumber
\eea
where $\eta(g) = \left[ \sum_{s_1} \frac{\beta(s_1g)}{\omega(g,s_1)\beta(s_1)} \right]$. It has been proved in Appendix.~B of Ref.~\cite{azses2020symmetry} that $f(g)$ is real, thus, it is clear that $Z_g$ is indeed real.

We now prove that $F_\calq \neq F_{\calq^{-1}}$ doubles the degeneracy between these families. In Ref.~\cite{azses2020symmetry} the following identity $Z_{g^{-1}}= \overline{Z_g}$ has been proven. This identity doubles the existing $\frac{|G|}{|H|}$ degeneracy. As $Z_g$ is real, we have the freedom to replace $\chi_\calq$ by its conjugate $\chi_{\calq^{-1}}$, while generating the same value $S_1(\calq) = S_1(\calq^{-1})$. As these two values are in different families, we get additional double degeneracy between these two families. Therefore, in the case $F_\calq \neq F_{\calq^{-1}}$ we have additional double degeneracy.

For the case $G = \intz_N \times \intz_N$ let us number the character by $\calq= (q_1,q_2)$. In this case, the group $H$ has elements that are the multiples of $\frac{N}{d}$ in both indices as can be proven from the framework of Ref.~\cite{azses2020symmetry}. For this group $G$ we have a $\intz_N$ SPT classification enumerated by an integer $m=0,1, \dots , N-1$. We define 
\be
d \equiv {\rm{{\rm gcd}}}(N,m).
\ee
We will next show that the signature is given by
\bea
\label{cases}
\begin{cases}
	\left( \underbrace{2 (N/d)^2}_{ \frac{d^2-1}{2} ~{\rm{times}}},(N/d)^2 \right),~~~d~~{\rm{odd}},   \\
	\left( \underbrace{2 (N/d)^2}_{ \frac{d^2-4}{2} ~{\rm{times}}},\underbrace{(N/d)^2}_{4 ~{\rm{times}}} \right),~~~d~~{\rm{even}}.
\end{cases}
\eea
Examples of such signatures can be found in Table I of Ref.~\cite{azses2020symmetry}.

We now prove Eq.~(\ref{cases}). It is clear that for $\calq=(0,0)$, which is the case that $\chi(g\in H)=1$, we have only one family, which we notate {\it{the homogeneous family}}, of $\frac{|G|}{|H|} =(\frac{N}{d})^2$ as all the entries are real. Moreover, we can explicitly write the $q$'s representing the characters in this family $q = (c_1 d,c_2 d)$, where $c_i$ is an integer ranging from $0$ to $\frac{N}{d}-1$. It is clear that all the others families are generated from {\it{the homogeneous family}} by adding $q'=(q'_1,q'_2)\neq 0$ with $q'_i<d$. We see that we have $d^2$ such families, each with $(\frac{N}{d})^2$ elements each. Thus, we classify the elements $q' \in \intz_{d}^2$ that equal to their inverse. In the case that $d$ is even, we have $4$ such possibilities $q'=(0,0),(0,\frac{d}{2}),(\frac{d}{2},0),(\frac{d}{2},\frac{d}{2})$, but in the case $d$ is odd, we have only $1$ such $q'=(0,0)$. Therefore, we have derived Eq.~(\ref{cases}).

\subsection*{FSPT cycling}
In the main text we showed that an FSPT with pumped charge $c$ induces cyclic evolution of symmetry sectors $\calq  \to \calq+c$. In the presence of static SPT order, degeneracies are present between sectors, which reduce the effect of the  cyclic evolution between blocks; in the fully degenerate case with a single family, or signature $(|G|)$, the Floquet operator has no effect at all. 

In order to describe the nontrivial cyclic evolution, we look at the cycling of the character families rather than individual blocks. We find that the FSPT with pumped charge $c$ induces a cyclic evolution of character families, which are defined by $q'$ effective charge. Generally, it takes the family $q'$ corresponding to a given $\calq ~{\rm{mod}}~ d$ to 
\be
\label{eq:mainresult}
q' \rightarrow q'+c \pmod d.
\ee
Hence, the pumped charge $c$ is recognized by its action on the degenerate family with the largest entanglement, and we notate its character family with $q'_{max}$ (defined mod $d$ on both indices). This family switches to $q'_{max} + c$ upon acting with $F$, thus, allowing us to recognize the charge based on the entanglement even in the degenerate case. However, this recognition is not complete as $c$ is recognized only mod $d$. Therefore, we see that even in the degenerate case we do have FSPT entanglement switching between degenerate blocks, opposed to element switching in the simplest $\intz_{N}$ case.

We exemplify the above paragraph with the example $N=4$ and $m=2$, which has $d=2$. For this case $q'=q'^{-1}$ for each $q'$. Let us explicitly write the 4 degenerate symmetry-sector families $F_{q'}$ of $S_1$, 
\bea
&&F_{0,0} = \{(0,0),(0,2),(2,0),(2,2)\}, \nonumber \\
&&F_{0,1} = \{(0,1),(0,3),(2,1),(2,3)\}, \nonumber \\
&&F_{1,0} = \{(1,0),(1,2),(3,0),(3,2)\}, \nonumber \\ 
&&F_{1,1} = \{(1,1),(1,3),(3,1),(3,3)\}. \nonumber
\eea
The FSPT with charge $\mathcal{Q}$ transforms these families such that $F_i \rightarrow F_{i+c}$. We notice that as $i$ is defined mod $d$, the FSPT charge effect on the families $F$ is invariant for equal charges mod $d$. These equivalent charges transform the families only within themselves and as all the sectors are degenerate the effect is unnoticeable. Thus, we have an interesting phenomenon of different charges leading to the same switching behavior; FSPT charges can be distinguished only modulo $d$ for the $m$'th SPT phase of $\intz_N \times \intz_N$.

The case $N=6$ and $m=3$ is more involved, as $d=3$ and hence we have several $q'$s such that $q'\neq q'^{-1}$. Let us write the base $F_{0,0}$ family explicitly for this case (note that there is additional degeneracy for $q'\neq q'^{-1}$ between $q'$ and $q'^{-1}$):
\bea
&&F_{0,0} = \{(0,0),(0,3),(3,0),(3,3)\}. \nonumber
\eea
We also notice that the values of $F_{q'} = F_{-q'}$, and hence we can sort the families into pairs of $q'$s:
\bea
&&\{(0,0)\} , \nonumber \\
&&\{(0,1),(0,2)) \}, \nonumber \\
&&\{(1,0),(2,0)) \}, \nonumber \\
&&\{(1,1),(2,2)) \}, \nonumber \\
&&\{(1,2),(2,1)) \}, \nonumber \\
&&\{(2,1),(1,2)) \}. \nonumber
\eea
Applying $F$ of the FSPT with charge $c$ we notice that the sectors shuffle in a discernable way. Specifically, assuming that $F_{0,0}$ holds all the entanglement, then, with each application of $F$ we transform this entanglement to another sector. As $F_{0,0}$ always exist, we see that we can always recognize the FSPT phase charge $c$ modulo $d$ even in this more involved case.

\end{document}